\documentclass[12pt]{article}
\usepackage[T1]{fontenc}
\usepackage{a4wide}
\usepackage{amsmath,amssymb,amsthm,epsfig}
\usepackage{graphics}
\usepackage{verbatim}

\newtheorem{lemm}{Lemma}

\begin{document}

\def\ad{\mathop{\rm ad}\nolimits}
\def\Ad{\mathop{\rm Ad}\nolimits}
\def\Det{\mathop{\rm Det}\nolimits}
\def\gl22{\mathop{\mathfrak{gl}(2|2)}\nolimits}
\def\un{\mathop{\mathfrak{u}(N)}\nolimits}
\def\eps{\epsilon}
\def\vareps{\varepsilon}

\def\Ker{\mathop{\rm Ker}\nolimits}
\def\tr{{\rm tr}}
\def\Tr{{\rm Tr}}
\def\mi{{\rm i}}
\def\i{{\rm i}}
\def\e{\rm e}
\def\sq2{\sqrt{2}}
\def\sqn{\sqrt{N}}
\def\defi{\stackrel{\rm def}{=}}
\def\t2{{\mathbb T}^2}
\def\T{\mathbb T}
\def\N{\mathbb N}
\def\Z{\mathbb Z}
\def\R{\mathbb R}
\def\C{\mathbb C}
\def\P{\mathcal{P}}
\def\L{\mathcal{L}}
\def\B{\mathcal{B}}
\def\O{\mathcal{O}}
\def\I{\mathcal{I}}
\def\tc{T_{\mathbb C}}
\def\s2{{\mathbb S}^2}
\def\hn{\mathcal{H}_{N}}
\newcommand {\Bl} {\big\langle}
\newcommand {\Br} {\big\rangle}

\newtheorem{prop}{Proposition}
\newtheorem{theo}{Theorem}
\newtheorem{conj}{Conjecture}

\title{Spectral properties of noisy classical and quantum propagators}

\author{St\'ephane Nonnenmacher\thanks{ Service de Physique Th\'eorique,
CEA/DSM/PhT Unit\'e de recherche associ\'ee au CNRS CEA/Saclay 91191
Gif-sur-Yvette c\'edex, France. e-mail:nonnen@spht.saclay.cea.fr}}


\maketitle

\begin{abstract}
We study classical and quantum maps on the torus phase space, in the presence of noise. 
We focus on the spectral properties of the noisy evolution operator, and prove that for any
amount of noise, the quantum spectrum converges to the classical one in the
semiclassical limit. The small-noise behaviour of the classical spectrum highly depends on the
dynamics generated by the map. For a chaotic dynamics, the outer spectrum consists in 
isolated eigenvalues (``resonances'') inside the unit circle, leading to an exponential
damping of correlations. On the opposite, in the case of a regular 
map, part of the spectrum accumulates along a one-dimensional ``string'' connecting 
the origin with unity, yielding a diffusive behaviour. 
We finally study the non-commutativity between the
semiclassical and small-noise limits, and illustrate this phenomenon by computing
(analytically and numerically) the classical and quantum spectra for some maps.

\end{abstract}

\section*{Introduction}

Numerical studies of chaotic dynamical systems inevitably face the problem of 
rounding errors due to finite computer precision. Indeed, the unstability of the dynamics
would require an infinite precision of the initial position, if one wants to compute the 
long-time evolution. Besides, any deterministic model describing the evolution
of some physical system is intrinsically an approximation, which neglects unknown but 
presumably small interactions with the ``environment''. A way to take this interaction into account
is to introduce some randomness in the deterministic equations, for instance through a term of
Langevin type. Such a term induces some diffusion, that is, 
some coarse-graining or smoothing of the phase space density. 
If the deterministic part of the dynamics is unstable, it has the opposite effect to
transform long-wave-length fluctuations of the density into short-wave-length ones, so the 
interplay between both components (deterministic vs. random) of the dynamics is a priori not obvious.
Rigorous results on the behaviour of noisy chaotic
systems have been obtained recently \cite{kifer,baladiyoung,BKL}: they
show that a usual property of deterministic chaos, namely the exponential mixing, 
still holds in presence
of noise. Actually, introducing some noise
provides a way to compute the exponential decay rate (and the subsequent ``resonances'') 
in a controlled fashion \cite{agam-blum,khodas,manderfeld,palla,ostruszka}.

In quantum mechanics, a (small) system is never perfectly isolated either, and is subject to
inevitable interaction with the environment, responsible for decoherence effects. The study
of decoherence has received a large attention recently, due to the (mostly theoretical) interest
in quantum computation \cite{chuang} and precise experiments measuring decoherence. In some
simple cases, one can study the dynamics of the full system (small plus environment), 
and obtain an effective dynamics of the small one \cite{caldeira-leggett}. Under suitable 
assumptions, this effective dynamics results in a memoryless
(super)operator acting on the quantum density of the small system.

A possible way to modelize this operator is by quantizing the ``Langevin term''
used in the classical
framework, that is introduce a random noise in the quantum evolution 
equation (for either continuous or discrete time dynamics). Averaging over the noise yields,  
in the continuous-time framework, a non-Hermitian ``Lindblad operator'' \cite{lindblad}, 
while in the discrete-time case the evolution can be cast into the 
 product of a unitary (deterministic) evolution operator with a ``quantum coarse-graining''
operator \cite{paz-saraceno}, this product being called from now on
the ``coarse-grained evolution operator''. 
Similar operators have also appeared in the theoretical 
study of spectral correlations for quantized maps \cite{det-det}, and in models of dissipative
quantum maps on the sphere \cite{braun}.

Several recent studies have been devoted to 
noisy or coarse-grained quantum maps \cite{khodas,manderfeld,paz-saraceno} in the semiclassical limit, 
emphasizing the respective roles of regular vs. chaotic regions in the classical phase space. 
The aim was either to study the time evolution of an initial density \cite{paz-saraceno},
or to compute the spectrum of the classical or quantum coarse-grained evolution operators 
\cite{khodas,manderfeld}. A different type of quantum dissipation operator, originating from
the theory of superradiance, was composed with a unitary quantum map on
the sphere, and studied in a series a papers by D.~Braun \cite{braun}. The author computed a 
Gutzwiller-like semiclassical trace formula for powers of the full (non-unitary) propagator, and showed their 
connection with traces of the corresponding 
classical dissipative propagator. 

In the present paper, we present some results about the spectrum of coarse-grained propagators, 
for maps defined on the 2-dimensional torus
(Section~\ref{s:classical perron-frobenius}). For the sake of simplicity, 
we restrict ourselves to either fully chaotic or fully regular maps.
In the limit of vanishing noise, the spectrum of the classical coarse-grained evolution 
operator behaves differently in these two (extreme) cases: for a chaotic map, the spectrum has a 
finite gap between unity and the next largest eigenvalue, 
due to exponential decay of correlations \cite{ruelle,BKL}, 
while in the regular case some
eigenvalues come arbitrary close to unity. To illustrate these results, 
we study in detail some linear systems (either chaotic or integrable), for which the eigenvalues 
may be computed analytically.

We then turn to the quantum version of these systems (Section~\ref{s:quantum prop}). After recalling
the setting of quantum maps on the
torus, we define the quantum coarse-graining operator, and then prove that
for any classical map and a fixed finite noise, the spectrum of the quantum coarse-grained 
evolution operator 
converges to the spectrum of the classical one in the semiclassical limit
(Theorem~\ref{th: semiclassical spectral stability}). 

We finally study in Section~\ref{s:non-commute}
the non-commutativity
of the semiclassical vs. vanishing-noise limits, using as examples 
the maps studied in the classical framework (Section~\ref{s:non-commute}). As a byproduct, 
we show that one can obtain the resonances of a classical hyperbolic map 
from the spectrum of an associated quantum operator (the quantum coarse-grained propagator), 
provided the coarse-graining is set to decrease slowly enough in the semiclassical limit
(we conjecture a sufficient condition for the speed of convergence). On the
opposite, for an integrable map the same limit yields a spectrum  densely filling 
one-dimensional curves (``strings'') 
in the unit disk, one of them containing unity as a limit point.


\section{Classical noisy evolution \label{s:classical perron-frobenius}}

The classical dynamical systems we will study are defined on a 2-dimensional
symplectic and Riemannian manifold, the torus $\t2 =\R^2/\Z^2$. The maps
are smooth ($C^\infty$), invertible and leave the symplectic form $dp\wedge dq$ (and
therefore the volume element $d^2x=dq\,dp$) invariant: they are canonical diffeomorphisms
of $\t2$.
Such a map $x=(q,p)\mapsto Mx$ naturally induces a \emph{Perron-Frobenius operator} 
$\P =\P_M$ acting on phase space densities $\rho(x)$: $[\P \rho](x)=\rho(M^{-1}x)$. 


\subsection{Spectral properties of the classical propagator\label{s:spectrum of class}}

In this section we review the spectral properties of the Perron-Frobenius operator
$\P_M$, depending on the map $M$ and on the functional space on which the operator
acts. The results presented are not new, but allow us to fix some notations.


\subsubsection{Spectrum on $L^2(\t2)$}

For any canonical
diffeomorphism $M$ and any $p\geq 1$, the spectrum of $\P_M$
on the Banach space $L^p(\t2)$ is a subset the unit circle and admits
$\rho_0(x)\equiv 1$ as invariant density. In particular, $\P $
is unitary on $L^2(\t2)$ and on its subspace $L^2_0(\t2)$ of 
zero-mean densities.

One can relate some dynamical properties of the map $M$
with the (unitary) spectrum of $\P $ on $L^2_0(\t2)$ \cite{cornfeld}.
For instance, if the dynamics $M$ leaves invariant a nonconstant observable 
$H(x)\in C^0(\t2)$ 
(e.g. if $M$ is the stroboscopic map of the Hamiltonian flow generated by $H$), 
then all observables of the type $\rho(x)=f(H(x))$ (with $f$ a smooth function) 
are invariant as well, so that the
eigenvalue $1$ of $\P$ is infinitely degenerate. The full spectrum of $\P$ will be explicitly
given for some integrable maps in Section \ref{s:integrable classical}. 

On the opposite, the map $M$ is ergodic iff $\P$
has no invariant density in $L^2_0$ (i.e. if $1$ is a simple eigenvalue in $L^2$). 
Stronger chaotic properties
may be defined in terms of the \emph{correlation function} between two densities 
$f,g\in L^2$:
\begin{equation}
\label{e:definition correlation}
C_{fg}(t)\defi \int_{\t2}dx\, f(x)g(M^{-t}x)=\left(f,\P^{t}g\right)
\end{equation}
(the time parameter $t$ will always take \emph{integer} values).
The map $M$ is said to be \emph{mixing} iff for any $f,g$, the correlation function behaves
for large times as:
\begin{equation}
\label{e:mixing}
C_{fg}(t)\stackrel{{|t|\to \infty }}{\longrightarrow }\int_{\t2}dx\, f(x)\int_{\t2}dx\, g(x).
\end{equation}
A slightly weaker property (weak mixing) is equivalent with the fact that $\P_M$ has
no eigenvalue in $L^2_0(\t2)$.


\subsubsection{Exponential mixing}
For a certain class of maps (e.g. the Anosov maps defined below), the convergence of mixing is 
exponentially fast, provided the
observables $f,g$ are smooth enough. One speaks of \emph{exponential
mixing} with a decay rate $\gamma>0$ if $|C_{fg}(t)-\int f\,dx\int g\,dx|\leq K_{fg}\e^{-\gamma |t|}$ 
for a certain constant $K_{fg}$.

This behaviour can sometimes be explained through the spectral
analysis of the Perron-Frobenius operator $\P$ acting on an {\it ad hoc} functional space 
$\B$, with $\B\not\subset L^2$ (in general $\B$ contains some distributions).
One proves that the operator $\P$ on $\B$ is {\it quasi-compact}: its spectrum consists of 
finitely many isolated
eigenvalues $\{\lambda_i\}$ (called \emph{Ruelle-Pollicott resonances}) situated in an annulus 
$\{r<\lambda_i<\lambda_0=1\}$ for some $r> 0$, plus possible essential spectrum inside the disk of
radius $r$.
In that case, assuming for simplicity that each $\lambda_i$ is a simple eigenvalue with spectral
projector $\Pi_i$, the 
spectral decomposition of $\P$ on $\B$ leads to the following asymptotic 
expansion for $C_{fg}(t)$ in the limit $t\to\infty$:
\begin{equation}
\label{e:correlation expansion}
C_{fg}(t)=\sum_i\lambda_i^t\, \langle f,\Pi_i g\rangle +\O \left(r^t\right).
\end{equation}
In the above expression, the brackets do not
refer to a Hermitian structure, but represent the integrals $\int f(x)[\Pi_i g](x)dx$. 
The first resonance $\lambda_0=1$ corresponds to the unique invariant density $\rho_0$, and
the decay rate (obviously independent of the observables $f,g$) is given by $\gamma=-\log|\lambda_1|$, 
resp. by $-\log r$ if there is no resonance other than unity. 

This type of spectrum was first put in evidence in the case of uniformly hyperbolic maps
by using Markov partitions to translate the dynamics on the
phase space into a simple symbolic
dynamics (namely a subshift of finite type) \cite{ruelle}. It was later extended to more general systems, 
including non-uniformly hyperbolic ones \cite{Baladi}. In the next sections, we will
introduce the Anosov diffeomorphisms on the torus, which often serve
as ``model'' for deterministic chaos.


\subsubsection{Anosov diffeomorphism on the two-dimensional torus\label{s:definition anosov}}

In this section we recall the definition and some properties of an Anosov diffeomorphism $M$
on the torus $\t2$ \cite{katok}. The Anosov property means that at each point $x\in \t2 $,
the tangent space $T_x\t2 $
splits into $T_x\t2 =E_x^s\oplus E_x^u$,
where $E_x^{u/s}$ are the local stable and unstable subspaces. The tangent
map $d_{x}M$ sends $E^{u/s}_x$ to $E^{u/s}_{M(x)}$, and there
exist constants $A>0,$ $0<\lambda_s<1<\lambda_u$ (independent
on $x$) s.t.
\begin{equation}
\label{e:Anosov}
\forall t\in \N ,\, \quad \Vert (d_{x}M^{t})_{|E^s_x}\Vert \leq A\lambda_s^t\quad  
\textrm{and }\quad \Vert (d_{x}M^{-t})_{|E^u_x}\Vert \leq A\lambda_{u}^{-t}.
\end{equation}
These inequalities describe the uniform hyperbolicity of $M$ on $\t2$. 
The splitting of $T_x\t2$ into $E^{u/s}_x$
has in general regularity $C^{1+\alpha}$ for some $1>\alpha>0$, meaning that it
is differentiable and its derivatives are H\"older-continuous with coefficient $\alpha$.

This uniform hyperbolicity implies that $M$ is ergodic and exponentially mixing w.r.to the Lebesgue measure 
(see next section).

Simple examples of Anosov diffeomorphisms are provided
by the linear hyperbolic automorphisms of $\t2$, defined by matrices $A\in SL(2,\Z )$
with $|\tr A|>2$. These maps are sometimes referred to as ``generalized
cat maps'', in reference to Arnold's ``cat map'' 
$A_{\textrm{Arnold}}=\left(\begin{array}{cc}
2 & 1\\
1 & 1
\end{array}\right) $ \cite{arnold}. 
We will study in some detail these linear maps and their quantizations, and
obtain a rather explicit description of the associated coarse-grained propagators.
One can perturb the linear hyperbolic automorphism $A$ with
the stroboscopic map $\varphi^1_H$ generated by some Hamiltonian $H(x)$ on the torus: 
$M\defi\varphi^1_H\circ A$. If the Hamiltonian $H$ is ``small enough'',
the perturbed map $M$ remains Anosov, and topologically conjugated with the linear map $A$.


\subsubsection{Ruelle resonances for Anosov diffeomorphisms\label{s:resonances on t2}}

As announced above, Anosov diffeomorphisms on $\t2$ are exponentially mixing, and their
correlation functions satisfy
expansions of the type \eqref{e:correlation expansion}. 
Although the original proofs made use of Markov
partitions \cite{ruelle}, we will rather describe a
more recent approach due to Blank, Keller and Liverani \cite{BKL},
which has the advantage to provide spectral results for the coarse-grained propagator as well. 

The strategy of \cite{BKL} is to define a Banach space $\B$ of densities on $\t2$ adapted to the map $M$: 
the space $\B$ contains 
distributions which are smooth along the unstable direction of $M$
but possibly singular (dual of smooth) along the stable direction. In particular, 
these functions are too singular to be in $L^2$, which allows a non-unitary 
spectrum of $\P$ acting on $\B$. 
The space $\B$ is defined in terms of an arbitrary parameter $0<\beta <1$, and 
satisfies
the continuous one-to-one embeddings: $C^1(\t2)\to \B \to C^1(\t2)^{*}$.

The space $\B$ depends on the map $M$, and
also on the direction of time: the space $\B_{M^{-1}}$ adapted to the map $M^{-1}$ is different from
$\B_{M}$. Although the map $M$ on the torus is invertible, the
dynamics of the operator $\P$ on the space $\B$ is \emph{irreversible}: its spectrum is qualitatively
very different from that of $\P^{-1}$. 

The authors indeed show that $\P$ is quasicompact in $\B$, with essential spectral radius
$r_{ess}$ bounded
above by $\sigma =\max (\lambda_u^{-1},\lambda_s^{\beta})$ (see
Eq.~(\ref{e:Anosov}) for the definition of $\lambda_{u/s}$). 
For any $1>r>\sigma $, the spectrum of $\P $
in the ring $R_r\defi\{r\leq |\lambda |\leq 1\}$ consists in isolated
eigenvalues, the Ruelle-Pollicott resonances $\{\lambda_i\}$. Therefore,
a spectral expansion similar with \eqref{e:correlation expansion} holds for
any pair of observables $f,g\in \B$, which includes in particular observables in $C^1(\t2)$
(the expansion might be slightly more complicated than \eqref{e:correlation expansion} due to 
possible finite degeneracies of the resonances).

A possible strategy to extend the results of
\cite{BKL} to maps and observables of regularity $C^k$ was discussed in the recent
preprint \cite{BaBa}.
Under stronger smoothness assumptions, namely for \emph{real-analytic} Anosov maps in two dimensions, 
H.~Rugh \cite{rugh} constructed a transfer operator
acting on observables real-analytic along the unstable direction, and ``dual of analytic'' along
the stable one; he showed that this operator is \emph{compact}, which means that
the essential spectral radius vanishes in that case.


\subsection{Coarse-grained classical propagator\label{s:classical coarse-gr propagator}}

In this section we precisely define the operator representing
the effect of ``noise'' on the deterministic evolution of $M$. This operator
is of diffusion type, it realizes a coarse-graining of the densities.
 

\subsubsection{Classical diffusion operator}
We consider a smooth probability density $K(x)$ on $\R^2$ 
with compact support. For simplicity, we also assume that $K(x)=K(-x)$.
From there, to any $\eps>0$ corresponds a probability density on the
torus, defined as 
$K_\eps(x)=\frac{1}{\eps^2}\sum _{n\in\Z^{2}}K\left(\frac{x+n}{\eps}\right)$.
Due to the compact support of $K$, we see that for small enough $\eps$ and any $x\in\t2$
this sum has at most one nonvanishing term (for $x$ close
to the origin). We define the coarse-graining operator $D_\eps$ on $L^2(\t2)$
as the following convolution: 
\begin{equation}
\label{e:classical coarse-g}
\forall f\in L^2(\t2),\quad [D_\eps f](y)=\int_{\t2}dx\, K_\eps(y-x)\, f(x).
\end{equation}
$D_\eps$ is a self-adjoint \emph{compact} operator on $L^2(\t2)$,
with discrete spectrum accumulating at the origin. 

We define the Fourier transform on the plane as 
$\tilde{K}(\xi)=\int_{\R^{2}}dx\, K(x)\e^{2\i\pi\xi\wedge x}$,
with the wedge product given by $\xi \wedge x=\xi_{1}p-\xi_{2}q$. From the assumptions
on the density $K$, the function
$\tilde{K}(\xi)$ is real, even and smooth. It takes
its maximum at $\xi =0$ (where it behaves as $\tilde{K}(\xi)=1-Q(\xi)+\O (|\xi|^{4})$,
with $Q(.)$ a positive definite quadratic form), and decreases fast for
$|\xi|\to \infty $. 

The plane waves (or Fourier modes) on $\t2$
are accordingly defined as $\rho_{k}(x)\defi \exp\left\{2\i\pi x\wedge k\right\}$,
for $k\in\Z^2$. They obviously form an orthonormal eigenbasis of the coarse-graining
operator:
\begin{equation}
\label{e:fourier=eigenstates}
\forall \eps >0,\:\forall k\in \Z^2,\quad D_\eps\rho_k=\tilde K(\eps k)\rho_k.
\end{equation}
The fast decrease of the eigenvalues as $|k|\to \infty$ implies that the
operator $D_\eps$ is not only compact, but also trace-class. 
It kills the small-wave-length modes, effectively truncating
the Fourier decomposition of $\rho(x)$ at a cutoff $|k|\sim\eps^{-1}$. For some
instances, we will actually replace the smooth function $\tilde K(\xi)$ by a sharp
cutoff $\Theta(1-|\xi|)$ (with $\Theta$ the Heaviside step function).


\subsubsection{Classical coarse-grained propagator}

The noisy dynamics associated with the map $M$ is represented by the product of 
the deterministic evolution $\P_M$ with the diffusion operator: 
\begin{equation}
\P_{M,\eps}\defi D_{\eps }\circ \P_M.
\end{equation}
It may be more natural to define the noisy propagator as the more `symmetric'
$D_\eps\circ\P_M\circ D_{\eps}$, but both definitions lead to the same 
spectral structure. 
$\P_{M,\eps}$ describes a Markov process defined by the deterministic map $M$
followed by a random jump on a scale $\eps$.

We now give some general properties of this operator, independent of the particular
map $M$. Like the regularizing operator $D_\eps$, $\P_{M,\eps}$ maps distributions
into smooth functions, and is \emph{compact} and 
\emph{trace-class} on any functional space containing $C^\infty(\t2)$ as a dense subspace, with a 
\emph{spectrum independent on the space}. Its
eigenvalues $\{\lambda_{\mu,\eps}\}_{\mu\geq 0}$ are inside the unit disk (only $\lambda_{0,\eps}=1$ is
exactly on the unit circle), they are of finite multiplicity
and admit the origin as only accumulation point. The eigenvalue $\lambda_{0,\eps}$ is simple, with
unique eigenfunction $\rho_{0}$. $\P_\eps$ maps a real density to a real density, therefore its spectrum
is symmetric with respect to the real axis.

In the next section we investigate in more detail the behaviour of these eigenvalues
in the limit of small noise, stressing the difference between chaotic vs. regular maps.


\section{Spectral properties of classical coarse-grained propagator
\label{s:spectrum classical}}

We describe more precisely the spectrum of $\P_{M,\eps}$, in the limit of small
noise, and for different classes of maps $M$. We start with the most chaotic maps, namely the
Anosov diffeomorphisms defined in Section~\ref{s:definition anosov}, the exponential mixing
of which was described in section~\ref{s:resonances on t2}. In a second part, we will then turn to the 
opposite case of ``regular'' maps on $\t2$.


\subsection{Anosov diffeomorphism\label{s:anosov}}
We use the same notations as in Section \ref{s:resonances on t2} for $M$ an Anosov map. 
It was proven
in \cite{BKL} that the spectrum of $\P_{M,\eps}$ outside some neighbourhood of the origin 
converges to the resonance spectrum of $\P_M$ on the Banach space $\B$. 

More precisely, for $M$ a smooth Anosov canonical diffeomorphism on $\t2 $ with foliations
of H\"older regularity $C^{1+\alpha }$ ($0<\alpha <1$),  one considers
the ``associated'' Banach space $\B =\B_M$ defined in terms of a coefficient $\beta<\alpha$, such
that the essential radius of $\P$ on $\B$ has for upper bound 
$\sigma =\max (\lambda_u^{-1},\lambda_s^\beta)$. 
The authors then construct 
a ``weak'' norm $\parallel .\parallel_w$ on $\mathcal{L}(\B)$, 
such that $\left\Vert \P_\eps-\P \right\Vert_w \to 0$ as $\eps\to 0$.
As a consequence, for any $1>r>\sigma$, any $\lambda$
in the annulus $R_r=\left\{ r\leq |\lambda |\leq 1\right\}$ and any
$\delta>0$ small enough, 
the spectral projector $\Pi^{(\eps)}_{B(\lambda,\delta)}$ of $\P_\eps$
(resp. $\Pi_{B(\lambda,\delta)}$ of $\P$) in the disk 
$B(\lambda,\delta)=\{z:\,|z-\lambda|\leq\delta\}$ satisfy 
$\left\Vert\Pi^{(\eps)}_{B(\lambda,\delta)}-\Pi_{B(\lambda,\delta)}\right\Vert_w
\stackrel{\eps\to 0}{\to} 0$.
Both projectors therefore have the same rank for $\eps$ small enough, this rank being
zero if the disk contains no resonance $\lambda_i$.

This proves that the spectrum
of $\P_\eps$ in the annulus $R_r$ converges (with multiplicity)
to the set of resonances $\left\{\lambda_i\right\}$ as $\eps\to 0$,
and the eigenmodes of $\P_\eps$ weakly converge to corresponding eigenmodes
of $\P$ (since the latter are genuine distributions, the convergence can only hold in a weak
sense). 

\subsubsection*{Remarks:}

\begin{itemize}
\item These results also apply to the `symmetric' coarse-graining $D_\eps\circ \P \circ D_\eps$.
\item It is reasonable to conjecture that these results hold as
well if the coarse-graining kernel $K(x)$ is not compactly supported on
$\R^2$, but decreases sufficiently fast, for instance if one takes the Gaussian 
$G(x)\defi \e ^{-\pi |x|^2}$,
$\tilde{G}(\xi)=\e^{-\pi |\xi|^2}$, as was done in
\cite{agam-blum,khodas}. As a broader generalization,
we will sometimes consider a coarse-graining defined by a sharp cut-off in Fourier
space, $\tilde K(\xi)=\Theta (1-|\xi|)$, similar to the method used
in \cite{manderfeld}; a finite-rank coarse-graining was also used in \cite{ostruszka} for 1-dimensional
noisy maps.
\item If the map $M$ and the kernel $K(x)$ are real-analytic, we conjecture that  
the eigenvalues of $\P_\eps$
on \emph{any} ring $R_r$ with $r>0$ converge to the resonances $\P$ in $R_r$ (cf. the remark
at the end of Section \ref{s:resonances on t2}).
\end{itemize}

The (discrete) spectrum of $\P_\eps$ is the same on any space $\mathcal{S}$ admitting 
$C^\infty$ as dense subspace, in particular on $L^2$, and this is the space we will consider from now on 
(more precisely, its subspace $L^2_0$).
This spectrum drastically
differs from the absolutely continuous unitary spectrum of the ``pure''
propagator $\P$ on that space (cf. Section \ref{s:spectrum of class}).
The unitary spectrum is thus \emph{unstable} upon the coarse-graining
$D_\eps$: when switching on the noise, the spectral radius of $\P_\eps$ on $L^2_0$
suddenly collapses from $1$ to $|\lambda_1|<1$.  As we will see
in the next sections, this collapse is characteristic of chaotic maps. We first
describe it explicitly for the case of hyperbolic linear automorphisms.


\subsubsection{Example of Anosov maps: the hyperbolic linear automorphisms 
\label{s:cat classical}}

In this section we review \cite{cornfeld,kifer} the spectral analysis of the (pure vs. noisy) propagator
when the map $A$ is a hyperbolic linear automorphism of $\t2$ (cf. Section \ref{s:definition anosov}). 
The unitary operator $\P_A$ on $L^2_0$ acts very simply on the basis of Fourier modes 
$\rho_k$, $k\in \Z_*^2=\Z^2\setminus 0$, namely as a permutation:
\begin{equation}
\label{e:permutation}
\left[\P_A\rho_k\right](x)=\rho_k(A^{-1}x)=\rho_{Ak}(x).
\end{equation}
This evolution induces \emph{orbits} on the Fourier lattice $\Z_*^2$, which we will denote
by $\O(k_0)=\{ A^t k_0,\: t\in \Z \}$. Due to the hyperbolicity
of $A$, each orbit is infinite, so that the
modes $\{\rho_{A^t k_0},\, t\in \Z \} $ span an infinite-dimensional
invariant subspace of $L^2_0$, which we call $Span\O(k_0)$.
The spectral measure of $\P_A$ associated with this subspace is of Lebesgue type (as usual
when the operator acts as a shift \cite{reed}). The number of distinct orbits being infinite,
the spectrum of $\P_A$ on $L^2_0$ is Lebesgue with infinite multiplicity.

Now we consider the coarse-grained propagator $\P_{\eps,A}=D_\eps\circ \P_A$.
Since the $\rho_k$ are eigenfunctions of $D_\eps$
(cf. Eq. (\ref{e:fourier=eigenstates})), $\P_{\eps,A}$ will also act
as a permutation inside each orbit $\O(k_0)$, but now at each step the
mode $\rho_k$ is multiplied by $\tilde{K}(\eps k)$. Therefore,
the operator $\P_{\eps,A}$ restricted
to the invariant subspace $Span\O (k_0)$ can be represented as follows
(the notation $\left( .,.\right) $ denotes the scalar product on $L^2_0$):
\begin{equation}
\label{e:infinite orbit}
\P_{\eps,A\mid \O(k_0)}=\sum _{t\in \Z }\tilde{K}(\eps A^{t+1}k_0)\, 
\left(\rho _{A^{t}k_0},\, .\, \right) \, \rho _{A^{t+1}k_0}.
\end{equation}
Since $|A^t k_0|\to \infty $ for $t\to \pm \infty $, the factors
$\tilde K(\eps A^{t+1}k_0)$ vanish in both limits. As a result, the
spectrum of $\P_{\eps,A\mid \O (k_0)}$ reduces to the single point $\{0\}$
(which is \emph{not} an eigenvalue, but essential spectrum) \cite{reed}. By taking all orbits
into account, the
spectrum of $\P_{\eps,A}$ on $L^2_0$ also reduces to $\{0\}$: for linear
hyperbolic maps,
the collapse of the Lebesgue unitary spectrum through coarse-graining is `maximal'.


\subsection{Coarse-graining of regular dynamics\label{s:integrable classical}}

In the previous sections we have considered classical propagators
of Anosov diffeomorphisms on $\t2$. We now describe the
opposite case of an \emph{integrable} map on $\t2 $. The notion of integrability
for a map is not so clear as for a Hamiltonian flow. In view of the examples below,
the definition should include the stroboscopic map $M_H=\varphi^1_H$ of the 
flow generated by an
autonomous Hamiltonian $H(x)$ on $\t2$, but it should also encompass
elliptic and parabolic automorphisms, as well as rational translations. A 
required property is that the phase space $\t2$ splits into a union of invariant 
1-dimensional (not necessarily connected) closed submanifolds, with possible ``critical energies''. 
Note that this condition excludes the (un)stable manifolds of an Anosov map, which are open.
A more or less equivalent condition for integrability is that the map leaves invariant a 
nowhere 
constant smooth function $H(x)$, the level curves of which provide the above submanifolds. 
As a result, any density $\rho(x)=f(H(x))$
is invariant through $\P_M$ as well, so that  $\P_M$
has an infinite-dimensional eigenspace of
invariant densities on $L^2_0$ (which we will call $V_{inv}$). 

The rest of the spectrum of $\P_M$ can be of various types, as we will see (it can be
pure point or absolutely continuous, be a mixture, etc..). For this reason, a general
statement concerning the coarse-grained spectrum of these maps cannot be very precise.

In the following sections we consider
some simple examples of integrable maps for which (part of) the spectrum can be analyzed in detail. 
We then discuss (mostly by hand-waving) the general case.


\subsubsection{Translations on the torus\label{s:classical translation}}

The simplest nontrivial maps on $\t2$ 
are the translations  
$x\mapsto T_v x=x+v\bmod \t2$, which do not derive from a Hamiltonian on the torus. 
A translation is either ergodic (yet non-mixing) or integrable
(see below).

The spectrum of the corresponding Perron-Frobenius operator $\P_v=\P_{T_v}$
on $L^2_0$ is easy to describe \cite{cornfeld}: $\P_v$ admits
the Fourier modes $\rho_k$ as eigenstates, with eigenvalues
$\e^{2\i\pi k\wedge v}$. The spectrum of $\P_v$ is therefore pure point, with possible 
degeneracies. The
spectrum of the coarse-grained propagator $P_{\eps ,v}=D_{\eps}\circ \P_v$
on $L^2_0$ is also easy to describe: each Fourier mode $\rho_k$
is an eigenfunction with eigenvalue $\tilde{K}(\eps k)\e^{2\i\pi k\wedge v}$
inside the unit disk. From the small-$\eps$ expansion 
$\tilde{K}(\eps k)\sim1-\eps^2 Q(k)$,
the eigenvalues corresponding to long wavelengths ($|k|\ll\eps^{-1}$) are
close to the unit circle for small $\eps$, while the short wavelength
eigenvalues ($|k|\gg\eps^{-1}$) accumulate near the origin.
We now describe how the global aspect of the spectrum qualitatively 
depends on the translation vector $v$.

\begin{itemize}
\item $T_v$ is ergodic iff the coefficients $v_1,\, v_2$ as well as
their ratio $\frac{v_1}{v_2}$ are irrational: in that case, the equation
$k\wedge v\in \Z $ has no solution for $k\neq 0$. The spectrum
of $\P_v$ forms a dense subgroup of the unit circle, all
eigenvalues being simple. In the limit $\eps\to 0^+$, the spectrum of $\P _{\eps ,v}$
becomes ``dense'' in the unit disk (a similar quantum spectrum is plotted on 
Fig.~\ref{fig:trans30-37-l37}, right).
\item If one of the coefficients $v_1,\, v_2,\, \frac{v_1}{v_2}$
is rational, $T_v$ leaves invariant a family of parallel 1-dimensional ``affine''
submanifolds, and is therefore integrable according to our definition. 
For instance, if we take $v_1=\frac{r}{s}$ (with
$r,s$ coprime integers) and $v_2$ irrational, then for any $q_0\in [0,1]$,
the union of vertical lines $\bigcup _{l=0}^{s-1}\left\{ q=q_{0}+\frac{l}{s}\right\}$
is invariant (if $s>1$, this set is non-connected). 
The spectrum of $\P_{v}$ is still dense on the circle,
but all eigenvalues are now infinitely degenerate: for any $k_0$, the
modes $k=k_0+(0,js)$, $j\in \Z $ share the
eigenvalue $\e^{2\i \pi k_{0}\wedge v}$. The eigenvalues
of $\P_{\eps ,v}$ are at most finitely-degenerate: to each phase $\e^{2\i \pi k_{0}\wedge v}$
corresponds a ``string'' of eigenvalues of decreasing moduli, the largest
one being at a distance $\sim \eps^{2}$ from the unit circle. As in the
previous case, the spectrum densely fills the unit disk as $\eps \to 0$.
\item If both $v_1,v_2$ are rational of the form $\frac{r_1}{s},\frac{r_2}{s}$
with $\gcd (r_1,r_2,s)=1$, each point of $\t2 $ is periodic with
period $s$, the map is integrable. The only eigenvalues of $\P_v$ are of
the phases $\e^{2\i \pi j/s}$, all being infinitely degenerate. The eigenvalues
of $\P_{\eps ,v}$ are at most finitely degenerate, they are grouped into
$s$ strings of phases $\e^{2\i \pi j/s}$, $j=0,\ldots ,s-1$.
For small $\eps$, the eigenvalues become dense along these strings, the
largest eigenvalue at a distance $\sim \eps^{2}$ from the unit circle (this spectrum
is similar with the quantum one plotted on Fig.~\ref{fig:trans30-37-l37}, left).
\end{itemize}
Comparing the first and second case, we see that the spectrum of $\P_\eps$ cannot 
unambiguously differentiate an ergodic from an integrable map: both may have 
eigenvalues close to the unit circle. 
On the other hand, the
second and third cases both correspond to integrable maps. Yet, their small-noise 
spectra look quite different from one another.


\subsubsection{Non-hyperbolic linear automorphisms of the torus\label{s:non-mixing automorphism}}

Another class of non-mixing linear maps on $\t2 $ is provided by the non-hyperbolic
linear automorphisms. These automorphisms split into two classes (for
notations, we refer to Section \ref{s:cat classical}):

\begin{itemize}
\item the elliptic transformations ($|\tr A|<2$), like for instance
the $\pi /2$-rotation
given by the matrix $J=\left( \begin{array}{cc}
0 & -1\\
1 & 0
\end{array}\right)$. 
As opposed to the hyperbolic case, each Fourier orbit 
$\O(k_0)=\{J^j k_0,\, j=0,\ldots,3\}$
is finite of period $4$, and $SpanO(k_0)$ splits into $4$ eigenspaces associated with the eigenvalues
$\{\i^l,\, l=0,\ldots ,3\}$.  Switching on  coarse-graining, the eigenvalues of $\P_{\eps,J}$ read
$\i^l\lambda_{k_0}$,
with $\lambda_{k_0}=\sqrt{|\tilde{K}(\eps k_0)\tilde{K}(\eps J k_0)|}$. The spectrum of 
$\P_{\eps,J}$ on $L^2_0$ therefore consists in four strings along
the four half-axes, which become dense in the limit $\eps \to 0$ (see fig. \ref{fig:dcas-four-140} for
the analogous quantum spectrum). Similarly, 
an elliptic transformation
of trace $\tr A=1$ will satisfy $A^6=1$, therefore the spectrum of 
$\P_{\eps,A}$ forms $6$ strings. 
An elliptic transformation of
trace $\tr A=-1$ will lead to $3$ strings.

\item the parabolic transformations (or parabolic shears), given by matrices of the
type $S=\left( \begin{array}{cc}
1 & s\\
0 & 1
\end{array}\right) $, $s\in \Z_*$. The dynamics reads $(q,p)\mapsto (q+sp,p)$,
so any periodic function of the momentum $p$ is a conserved quantity. 
The Fourier vector $k_0=(k_1,k_2)$ generates the 
orbit $\O (k_0)=\left\{ k_0+(jsk_2,0),\, j\in \Z \right\} $: 
if $k_2=0$, the mode $\rho_{k_0}$
is invariant and the orbit is a singleton; on the opposite, if $k_2\neq 0$,
$\O(k_0)$ is infinite, and leads to Lebesgue spectrum for
$\P_{S\mid \O(k_0)}$. The full spectrum of $\P_S$ on $L^2_0$
is therefore the union of the infinitely degenerate eigenvalue $1$ with
a countable Lebesgue spectrum \cite{cornfeld}. In the case $k_2\neq 0$,
the noisy propagator $\P_{\eps,S}$ acts on $Span\O (k_0)$ as
in Eq.~(\ref{e:infinite orbit}), upon replacing $A$
by $S$. Since the wavevectors $|S^t k_0|$ diverge in both limits
$t\to \pm \infty $, the associated spectrum reduces to the singleton $\{0\}$. 
On the opposite, each mode $\rho_{(l,0)}$ is eigenstate
of $\P_{\eps,S}$ with eigenvalue $\tilde K\left( \eps (l,0)\right)$:
these modes form a string along the real axis, which becomes dense as $\eps \to 0$.
\end{itemize}


\subsubsection{Nonlinear shear\label{s:nonlinear shear classical}}

If we perturb the linear parabolic shear $S$ (cf. last section)
with the stroboscopic
map of the flow generated by a Hamiltonian of the form $F(p)$, we obtain a
nonlinear shear $(q,p)\mapsto (q+sp+F'(p),p)$.
We assume that $s>0$, and that the Hamiltonian $F$
satisfies $s+F''(p)>0$ everywhere. This map still conserves
momentum, so that any density $\rho(p)$ is invariant. Among
these, the Fourier modes $\rho_{(l,0)}$ are eigenstates of $\P_\eps$, with 
eigenvalues $\tilde K(\eps (l,0))$:
they form the same string of real eigenvalues as for the linear
shear. 

For any $n\in \Z_*$,
$\P$ and $\P_\eps$ leave invariant the subspace 
$V_n=Span\left\{\e^{2\i \pi nq}\rho(p),\: \rho \in L^2(\mathbb{T}^1)\right\}$,
and their spectra on this subspace can be partially described, at least if one
takes $F'$ small enough and replaces the kernel $\tilde K(\xi)$
by the sharp cutoff $\Theta (1-|\xi_1|)\Theta (1-|\xi_2|)$ (such that $D_\eps$ projects
to a finite-dimensional subspace). The spectrum
of $\P$ on $V_n$ is absolutely continuous, while that of $\P_\eps$
is contained in a small neighbourhood of the origin, uniformly with
respect to $n$ and $\eps $ (see Appendix \ref{a:toeplitz} for details).
The spectrum should be qualitatively the same if $\tilde{K}(\xi)$ is of 
fast decrease at infinity.


\subsubsection{Stroboscopic map of the Harper Hamiltonian flow\label{s: Harper classical}}

As a last example of integrable map, we consider the Harper Hamiltonian 
on the torus 
$$
H(x)=\cos(q)+\cos(p)=\frac{1}{2}(\rho_{(1,0)}+\rho_{(-1,0)}+\rho_{(0,1)}+\rho_{(0,-1)}),
$$
and take its stroboscopic map $M=\varphi^1_H$. 
The invariant densities are of the form $\rho(x)=f(H(x))$. As opposed
to what happened for the parabolic shear, $V_{inv}$ is not invariant
through the diffusion operator $D_\eps$, so the spectrum of $\P_\eps$ 
is more complicated
to analyze. Still, if we take a coarse-graining kernel satisfying
$\tilde{K}(\xi)=\tilde{k}\left( |\xi_1|+|\xi_2|\right)$ along the
axes of slopes $0,\,\pm 1/2,\,\pm 1,\,\pm 2,\,\infty$) 
then the invariant functions 
$H(x)$, $\left( H(x)\right)^2-1$
and $\left( H(x)\right)^3 -\frac{9}{4}H(x)$ are eigenstates of $D_\eps$
and therefore of $\P_{\eps}$, with respective eigenvalues $\tilde{k}(\eps )$,
$\tilde{k}(2\eps )$ and $\tilde{k}(3\eps)$ (if the kernel only satisfies
$\tilde K((0,\zeta))=\tilde K((\zeta,0))$ together with even parity, then $H(x)$ will
be eigenstate of $\P_\eps$ with eigenvalue $\tilde K((\eps,0))$).
These eigenvalues are real and approach unity as $\eps \to 0$. Numerically,
they are the three first elements (resp. the first element) of a string of real
eigenvalues connecting unity to the origin, the further elements of the string
being mixtures of invariant and non-invariant densities (see Fig.~\ref{fig:snlin-harper40},
right for a similar quantum spectrum).


\subsubsection{General behaviour for an integrable map}

After these examples, we want to describe the spectrum of $\P_{M,\eps }$
for $M$ an integrable map, say the stroboscopic map of an autonomous Hamiltonian
$H$. As we already explained, the space $V_{inv}$ of invariant densities
is infinite-dimensional, containing all densities $f(H(x))$. When
applying the coarse-graining $D_\eps$, one generally mixes these invariant
densities with non-invariant ones (as opposed to what happens for the linear
maps described above). Using ideas of degenerate perturbation theory,
we conjecture the following behaviour, which is supported by numerical investigations
\cite{manderfeld,khodas}.

The invariant subspace $V_{inv}$ contains modes $\rho_{lw}$ with
long-wavelength fluctuations (that is, densities for which the Fourier spectrum is
concentrated in a finite region near $k=0$). For $\eps$ small, these modes
are hardly modified by the coarse-graining operator: $D_\eps\rho_{lw}-\rho_{lw}=\O(\eps^2)$.
This suggests that $\P_\eps$ has an eigenstate of the type $\rho_{lw}+\O (\eps^2)$,
of eigenvalue $1-\O(\eps^2)$, this eigenstate being close to an invariant state of 
$\P$. On the opposite, $V_{inv}$ also
contains highly fluctuating modes, which will be very damped by $D_\eps$,
and should lead to eigenstates of $\P_\eps$ close to the origin. As
a whole, the hybridization of invariant modes with noninvariant ones leads to a string of 
eigenvalues connecting unity to the origin, this string being symmetrical w.r.to the real axis. 
Notice that the projected operator $\Pi_{inv}\P_\eps\Pi_{inv}=\Pi_{inv}D_\eps\Pi_{inv}=$ (where
$\Pi_{inv}$ is the orthogonal projector on the space $V_{inv}$) is self-adjoint, therefore its spectrum
can be estimated through the min-max method: for $\eps\to 0$ it consists in a
dense string of eigenvalues between unity and zero. I believe that the largest eigenstates (-values)
of this projected operator are close to eigenstates (-values) of $\P_\eps$.


\subsection*{Conclusion: qualitatively different spectra}

We have exhibited a \emph{qualitative difference} of the coarse-grained Perron-Frobenius
spectra between, on one side, the chaotic maps (Anosov), on the other side,
non-mixing maps, including ergodic translations and integrable maps. 

In the latter case, the spectrum
of $\P_\eps$ on the subspace $L^2_0(\t2)$ comes close to the unit circle
for small $\eps$, either along 1-dimensional strings ``connecting''
the origin to infinitely degenerate eigenvalues of $\P$ (among which unity), or by densely
filling the unit disk (for ergodic translations). A common feature is that
any annulus $\{R<|\lambda|<1\}$ (or any open neighbourhood of unity)
contains more and more eigenvalues of $\P_\eps$ in the small-noise
limit $\eps\to 0$. 

On the opposite, for an Anosov
map the spectrum of $\P_\eps$ on $L^2_0$ is contained inside a disk
of radius $R<1$, uniformly for small enough $\eps$. This spectrum 
consists in
a finite number (possibly zero) of finitely degenerate eigenvalues (asymptotically 
close to the Ruelle resonances) 
inside an annulus $\{ r<|\lambda_{\mu,\eps} |\leq R\}$, the remaining eigenvalues having
moduli smaller than $r$.

The numerical results of \cite{manderfeld,khodas} go beyond this statement:
the authors consider systems with \emph{mixed dynamics} (not to be mistaken with the
``mixing'' property of Anosov maps), 
that is systems for which the phase space splits into ``regular
islands'' embedded in a ``chaotic sea''. They study the spectrum of the coarse-grained
Perron-Frobenius operator (the coarse-graining is taken as a cut-off in Fourier
space), and show the presence of eigenvalues close to the unit circle,
the eigendensities of which are supported on the regular islands; on the other hand,
they also find some eigenvalues inside the unit circle, which are associated
with the chaotic part of phase space, and therefore interpreted as (generalized)
Ruelle resonances. 
Yet, the nature of the ``chaotic sea'' in such systems is not well
understood at the mathematical level, so that a rigorous spectral 
analysis of the propagator seems a quite distant goal.


\section{Quantum coarse-grained evolution\label{s:quantum prop}}

After studying the effect of noise on classical propagators, we turn to noisy
quantum maps, obtained by quantizing the classical ones. Although we will 
restrict ourselves to
maps on the 2-torus,  the main result of this
section (Theorem \ref{th: semiclassical spectral stability}) can be straightforwardly
generalized to quantum maps on any compact phase space, provided one adapts the
definition of the coarse-graining operator (see the discussion in 
Section~\ref{generalization}).
We start by recalling the setting of quantized maps on the 2-torus.


\subsection{Quantum propagator on the torus}

\subsubsection{Quantum Hilbert space and observables\label{s:weyl quantization}}

We shortly review the construction of quantum mechanics on $\t2 $, in order
to fix notations. 
For any value of $\hbar>0$, the Weyl-Heisenberg group associates
to each vector $v=(v_1,v_2)\in \R^2$ the ``quantum
translation'' 
$\hat T_v=\exp\left\{\frac{\i}{\hbar}(v_2\hat q - v_1\hat p)\right\}$
which acts unitarily on $L^2(\R)$.
These translations satisfy the group relations 
$\hat T_v\hat T_{v'}=\e^{\frac{-\i}{2\hbar}v\wedge v'}\, \hat T_{v+v'}$.

The ``torus wavefunctions'' are then defined as
distributions $|\psi\rangle \in \mathcal{S}'(\R)$ satisfying the periodicity conditions
$\hat T_{(0,1)}|\psi\rangle =\hat T_{(1,0)}|\psi\rangle =|\psi\rangle $.
Due to the group relations, such distributions exist iff $\hbar$
satisfies the condition $(2\pi\hbar)^{-1}=N\in\N$ (such a value of $\hbar $ 
will be called \emph{admissible}). In that case they form a space of dimension $N$, 
which will be noted
$\hn $ \cite{hannay}. A basis of this space is provided by the ``Dirac combs''
$\left\{|q_j\rangle_N\right\}_{j=0,\ldots ,N-1}$ defined as:
\begin{equation}
\label{e:position basis}
\langle q|q_j\rangle_N=\sum _{\nu \in \Z }\delta (q-q_j-\nu),
\quad \textrm{with}\quad q_j=\frac{j}{N}.
\end{equation}
 By construction, $|q_j\rangle =|q_{j+N}\rangle$, so the index $j$ must be understood
modulo $N$. For practical reasons, we will choose the following
representatives of integers modulo $N$: 
$\Z_N=\left\{-\frac{N}{2}+1,\ldots ,\frac{N}{2}\right\}$
for $N$ even, $\Z_N=\left\{ -\frac{N-1}{2}+1,\ldots ,\frac{N-1}{2}\right\}$
for $N$ odd. 

The quantum translation $\hat T_v$ acts inside $\hn$
iff $v$ is on the square lattice of side $1/N$, that is 
$v=\left(\frac{V_1}{N},\frac{V_2}{N}\right) $
with $V_i\in \Z $. Besides, each translation with integer coefficients acts on
$\hn$ as a multiple of the identity. As a result, the set
$\left\{\hat T_{V/N},\: V\in \Z_N^2\right\}$
forms a basis of the space of linear operators on $\hn$, denoted by $\mathcal{L}(\hn)$.
This basis can be used to define the Weyl quantization of smooth observables 
$f\in C^\infty(\t2 )$.
From the Fourier decomposition $f=\sum_{k\in\Z^2}\, \tilde f(k)\rho_k$,
the Weyl quantization of the observable $f$ is defined as
\begin{equation}
\label{e:quantum fourier}
\hat f=Op_N(f)\defi \sum_{k\in\Z^2}\tilde f(k) \, \hat T_{k/N}
=\sum_{k\in\Z_N^2}\hat T_{k/N}\, 
\left(\sum_{\nu \in \Z^2}(-1)^{N\nu_1\nu_2+k\wedge\nu}\: 
\tilde f\left( k+N\nu \right) \right) .
\end{equation}
The ``converse'' of Weyl quantization, that is the \textbf{Weyl symbol} (or Wigner
function) $W_{\hat B}(x)$ of an operator
$\hat B\in \mathcal{L}(\hn )$ is also easily defined through its Fourier transform: 
for each $k\in \Z^2$, 
its Fourier coefficient $\tilde W_{\hat B}(k)$ is given by
$\tilde W_{\hat B}(k)=\frac{1}{N}\tr \left(\hat T^\dagger_{k/N}\hat B\right)$.
These Fourier coefficients
are $N$-periodic (up to a sign), so that the ``function'' $W_{\hat B}(x)$
is a periodic combination of Dirac peaks on the lattice of side 
$1/2N$ \cite{hannay}. Alternatively,
a \textbf{polynomial Weyl symbol} was defined in \cite{DEGI} as the finite
sum 
$$
W^P_{\hat B}(x)=\sum_{k\in \Z^2_N}\tilde W_{\hat B}(k)\e^{2\i \pi x\wedge k}.
$$
 As opposed to the former symbol, the polynomial symbol depends on the specific choice for
the representative $\Z_N$ (the choice we made, with maximum symmetry
around the origin, seems more natural in this respect). The polynomial symbol map
together with $Op_N$ yield an \emph{isometric isomorphism} between the subspace
$\I_N\defi Span\left\{ \rho_k,\: k\in \Z_N^2\right\} $ of $L^2(\t2 )$
and the space of observables on $\hn$ equipped with the Hilbert-Schmidt
scalar product $(\hat B,\hat C)=\frac{1}{N}\tr (\hat B^\dagger\hat C)$.
Since the Hilbert-Schmidt norm differs from the usual operator norm on 
$\mathcal{L}(\hn)$,  we will 
denote by $L^2_N$ the space of observables on $\hn$ (that is, $N\times N$ matrices)
equipped with the Hilbert-Schmidt norm.


\subsubsection{Quantization of canonical maps \label{s:quantize maps}}

We briefly explain how one quantizes a canonical
map $M$ on $\t2$. The aim is to define for each $N\in\N$ a unitary operator 
on $\hn$ (which we will denote by $\hat M_N$ or simply $\hat M$) which satisfies
prescribed semiclassical properties \cite{tabor}. These
properties do not unambiguously define the sequence of unitary matrices,
so a choice has to be made (in \cite{zelditch}, a Toeplitz quantization is proposed for
symplectic maps on K\"ahler manifolds). 
We present here another quantization prescription, which uses the following
decomposition of any canonical map $M$ on $\t2$ \cite{CZ}:
\begin{equation}
\label{e:classical decompo}
M=A\circ T_{v}\circ \varphi^{1}_{H}.
\end{equation}
In this formula, the linear automorphism $A\in SL(2,\Z)$ and the translation
$T_v$ are uniquely defined. On the opposite, the last factor correspond to the stroboscopic
map of the flow of a time-dependent periodic Hamiltonian $H\in C^{\infty }(\t2_x\times\T_t)$;
this Hamiltonian is not unique, since two Hamiltonians $H_1\neq H_2$ may
lead to the same stroboscopic map $\varphi^1_{H_1}=\varphi^1_{H_2}$.
From this decomposition, on can quantize $M$ as follows \cite{KMR}. First, one quantizes
\emph{à la} Weyl the Hamiltonian $H(x,t)$ into the operator $\hat H(t)$ on $\hn$,
then take for the quantization of $\varphi^1_H$ the time-ordered exponential
$\mathcal{T}\e^{-\i \int_0^1 dt \,\hat H(t)/\hbar }$. Second, one may
quantize the translation $T_v$ with the quantum translation $\hat T_{v^{(N)}}$, where
the vector $v^{(N)}$ belongs to the $1/N$-lattice
and is close to $v$, for instance take 
$v^{(N)}=\frac{\left( \left[N v_1\right],\left[N v_2\right]\right)}{N}$
\cite{marklof}.
Third, provided $A$ satisfies the condition $A\equiv \textrm{Id}_2\bmod 2$ or
$A\equiv \sigma_x\bmod 2$, the linear automorphism $A$ is ``naturally'' 
quantized into a unitary matrix
$\hat A_N$ \cite{hannay} (this condition may be relaxed if one
generalizes the quantum Hilbert space $\hn$ to arbitrary ``Bloch angles'' \cite{BDB}). 
Finally, the map $M$ is quantized on
$\hn $ as the product:
\begin{equation}
\label{e:quantum decompo}
\hat M_N=\hat A_N\circ \hat{T}_{v^{(N)}}\circ 
\mathcal{T}\e ^{-\i 2\pi N\, \int_0^1 dt\,\hat H(t)}.
\end{equation}
This choice for the quantum map automatically satisfies the Egorov property 
\cite{BDB,marklof}: 
for any classical observable $\rho\in C^{\infty}(\t2)$, 
\begin{equation}
\label{e:Egorov}
\hat M_N\, Op_N(\rho)\, \hat M_N^{-1} - Op_N(\rho\circ M^{-1})
\stackrel{N\to \infty }{\longrightarrow }0.
\end{equation}
This means that the quantization and finite-time evolution of densities commute 
in the semiclassical limit (the convergence holds for the operator
norm in $\mathcal{L}(\hn)$). 

In practice, the classical maps we consider are all defined as products of
automorphisms, translations and Hamiltonian maps, so they admit a ``natural'' quantization.


\subsubsection{Propagator of quantum densities}

The quantum map $\hat M$ propagates quantum states $|\psi \rangle \in \hn $. 
A density matrix $\hat\rho$ is an element of
the space $L^2_N=\hn \otimes \hn ^*$, and its evolution
through the quantum map reads $\hat\rho\mapsto \hat M\hat\rho\hat M^{-1}
=\ad (\hat M)\hat\rho$.
The operator $\hat\P_M\defi \ad (\hat M)$ acting on the space $L^2_N$
is the quantum analogue of the Perron-Frobenius operator $\P_M$ acting on classical
densities. $\hat\P_M$
is unitary, with eigenvalues $\{\e^{\i (\theta_j-\theta_i)};\, i,j=1,\ldots,N\}$,
where $\{\e^{\i \theta_j}\}$ are the eigenvalues
of the matrix $\hat M$. As a contrast with its classical analogue, the
space of invariant densities through $\hat\P_M$ is at least $N$-dimensional,
since it contains all the rank-1 projectors $|\phi_i\rangle \langle \phi_i|$,
where $|\phi_i\rangle $ are the eigenstates of $\hat M$. 

The operator $\hat\P$ acting on densities is called a \emph{superoperator}, 
to contrast with an operator acting on $\hn $. Being the adjoint action of
a unitary matrix, it conserves the \emph{purity} of the density, which means
that a pure density $\hat\rho=|\psi \rangle \langle \psi |$ is mapped
onto a pure density $\hat\P\hat\rho=|\psi '\rangle \langle \psi '|$.
We notice that $\hat\P$ conserves the \emph{trace} of the density, that
is the quantum counterpart of the total probability of the density $\rho$.


\subsection{Quantum coarse-graining operator}

As we remarked above, the spectrum of $\hat\P$ on $L^2_N$
is qualitatively different from that of $\P$ on $L^2(\t2)$:
the former has at least $N$ invariant eigenstates, while the latter may
have only one if the map $M$ is ergodic. In Section~\ref{s:spectrum of class}, 
we explained how
the spectrum of $\P$ was sensitive to the functional space on which $\P$ acts.
This is not any more the case in the quantum framework, since the propagator is 
a finite-dimensional matrix.
 Still, we showed in Section~\ref{s:spectrum classical} 
an alternative way to obtain the (nonunitary) resonance spectrum of an Anosov map, namely
by introducing some noise and studying the 
spectrum of the noisy propagator $\P_{\eps}=D_{\eps}\circ\P$ on the
space $L^2(\t2)$. This procedure can also be carried out at the quantum level, by
first  defining a ``quantum coarse-graining'' or ``quantum diffusion'' 
(super)operator
$\hat{D}_{\eps}$, use it to construct a quantum coarse-grained propagator $\hat{\P}_{\eps}$,
then study the spectrum of the latter on $L^2_N$. 
To connect the classical and quantum 
frameworks, we will consider the semiclassical limit $N\to\infty$.


\subsubsection{Definition}
We define the coarse-graining superoperator by analogy with the classical
one (Eq.~(\ref{e:classical coarse-g})). The latter can be expressed as
follows: 
\begin{equation}
\label{e:classical c-g2}
\forall f\in L^2(\t2 ),\quad \left( D_{\eps }f\right) (x)
=\int_{\t2}dv\, K_{\eps }(v)\, \rho (x-v)
=\int_{\t2}dv\, K_{\eps }(v)\, \left( \P_{v}\rho \right) (x),
\end{equation}
 where $\P_v$ is the Perron-Frobenius operator for
the translation $T_{v}$. Since  $T_v$ is quantized on $\hn$
into the unitary matrix $\hat{T}_{v^{(N)}}$, a natural way to define a
quantum coarse-graining superoperator would be through the integral
$$
\int_{\t2}dv\, K_{\eps }(v)\, \hat{\P }_{v^{(N)}}.
$$
For convenience, we prefer a slightly different definition. The map $v\mapsto v^{(N)}$
is constant on squares of side $1/N$, so the above integral reduces to
a finite sum over $V\in\Z_N^2$ , with each operator $\hat{\P}_{V/N}$ multiplied by the
average of $K_{\eps}$ over the corresponding square. $K_{\eps}$
being a smooth function, this average is semiclassically close to the value
$K_{\eps }(V/N)$, therefore for $N$ large the above integral is well approximated by the
sum:
\begin{equation}
\label{e:quantum c-g}
\hat{D}_{\eps }=\frac{F(\eps ,N)}{N^{2}}
\sum _{V\in \Z _{N}^{2}}K_{\eps }(V/N)\, \ad \left( \hat{T}_{V/N}\right) .
\end{equation}
The prefactor $F(\eps,N)$ is needed to guarantee that $\hat{D}_{\eps}$
conserves the trace, that is
$\hat{D}_{\eps }\hat{\rho}_0=\hat{\rho}_0$.
This factor is easily expressed in terms of the 2-dimensional theta
function:
\begin{equation}
\label{e:theta_k}
\theta _{K}(\eps N,\zeta )\defi 
\sum _{\nu \in \Z ^{2}}\tilde{K}\left( \eps N(\nu +\zeta )\right) ,\quad \zeta \in \t2 .
\end{equation}
For Gaussian coarse-graining $K(x)=G(x)=\e ^{-\pi |x|^{2}}$, this function
reduces to the product of two classical Jacobi 1-dimensional theta functions. 
In the limit $\eps N\gg 1 $, the function
$\theta_{K}$ is peaked around the point $\zeta =0$, due to
the fast decrease of $\tilde{K}(\xi )$. Now, one easily checks that 
$F(\eps ,N)=\theta_{K}(\eps N,0)^{-1}$ converges to $\tilde{K}(0)=1$ in the limit
$\eps N\to \infty$. 


\subsubsection{Spectrum of the quantum coarse-graining operator}

The spectrum of $\hat{D}_{\eps }$ on $L^{2}_{N}$ is as easy to analyze
as that of $D_{\eps}$ (see Eq.~(\ref{e:fourier=eigenstates})). Using
the group relation 
\begin{equation}
\label{e:translation group}
\ad (\hat{T}_{v})\hat{T}_{v'}=\e ^{-\i \frac{v\wedge v'}{\hbar }}\hat{T}_{v'},
\end{equation}
we see that for any $k\in \Z ^{2}_{N}$, the quantum translation $\hat{T}_{k/N}$
($=Op_{N}(\rho_k)$) is eigenstate of $\hat{D}_{\eps}$
with eigenvalue 
\begin{equation}
\label{e:d_eps,k}
d^{(N)}_{\eps ,k}=\frac{F(\eps ,N)}{N^{2}}\sum _{V\in \Z _{N}^{2}}K_{\eps }(V/N)
\e^{2\i \pi k\wedge \frac{V}{N}}=\frac{\theta _{K}(\eps N,k/N)}{\theta _{K}(\eps N,0)}.
\end{equation}
For fixed $k$ and $\eps N\to\infty$, one has the asymptotics
$d^{(N)}_{\eps ,k}=\tilde{K}(\eps k)+\O((\eps N)^{-\alpha })$ for any
power $\alpha >0$, so that the eigenvalue of $\hat{D}_{\eps}$ associated with
$\hat{T}_{k/N}$
converges to the eigenvalue of$D_{\eps}$ associated with its symbol $\rho_k$. 
The estimate is sharper in the Gaussian case:
\begin{equation}
\label{e:quantum Gaussian case}
\forall k\in \Z _{N}^{2},\quad d^{(N)}_{\eps ,k}
=\e ^{-\pi \eps ^{2}k^{2}}+\O (\e ^{-\frac{\pi (\eps N)^{2}}{4}}),
\end{equation}
so that the relative deviations between classical and quantum eigenvalues are
exponentially small, uniformly on the modes 
$k\in \left\{ |k_{1}|,\, |k_{2}|\leq \frac{N}{2}(1-\delta )\right\}$
(with $\delta >0$ fixed). For the Gaussian case, both classical and quantum spectra
are strictly positive, which is not true in general. The spectrum of
$\hat{D}_{\eps}$ for a Gaussian noise is plotted in Fig.~\ref{fig:dcas-four-140} (left).

If we replace $\tilde{K}(\xi)$ by the sharp cutoff $\Theta (1-|\xi |),$
then $d^{(N)}_{\eps ,k}=1$ for $|k|<\eps ^{-1}$, $d^{(N)}_{\eps ,k}=0$
otherwise: the coarse-graining truncates the expansion (\ref{e:quantum fourier}),
keeping only short wavevectors (modulo $N$). 
In other words, $\hat{D}_{\eps }$ truncates the Fourier series of
the polynomial Weyl symbol $W^{P}_{\hat{\rho}}$, keeping only the coefficients
$|k|<\eps^{-1}$.

A Fourier cutoff was also used as definition for the quantum coarse-graining 
in \cite{manderfeld},
but there the cutoff was applied to the \textbf{Husimi function} of $\hat{\rho}$ instead
of its polynomial Weyl symbol.


\subsubsection{Probabilistic interpretation and generalizations of coarse-graining}
\label{generalization}
In a different physical framework (one-dimensional quantum spin chains), Prosen
recently defined a similar quantum coarse-graining by truncating the densities on
finite-dimensional spaces \cite{prosen2}. In this case, the quantum densities
are decomposed into sums of spin operators acting on finite sequences of spins
(e.g. $\sigma_1(x)\sigma_1(x+1)\cdots \sigma_1(x+l)$). The truncation
consists in keeping only those operators for which $l\leq \eps^{-1}$.

The RHS of Eq.~(\ref{e:quantum c-g}) expresses the superoperator $\hat{D}_\eps$ 
in the \emph{Kraus representation}, {i.e.} as a sum $\sum_j\ad (\hat{E}_j)$, 
where the operators $\hat{E}_j$ on $\hn $ satisfy the condition 
$\sum_j\hat{E}_j^\dagger\hat{E}_j=\textrm{Id}_N$. Kraus superoperators
conserve  the trace and the ``complete positivity'' of density matrices
\cite{chuang}. 

The RHS of Eqs.~(\ref{e:classical c-g2},\ref{e:quantum c-g}) may be interpreted
as a \emph{random global jump} for the density, which jumps at a distance $v$
with a probability $\propto K_\eps(v)$. $D_\eps\rho$ and its
quantum counterpart represent the average over all these random jumps. Since
$D_\eps$ or $\hat{D}_\eps$ do not depend on time, they therefore
represent the classical and quantum versions of a memoryless Markov process. 

In a different scope, a superoperator similar to $\hat{D}_\eps$ was
used in \cite{det-det} to study the spectral correlations of the quantum map
$\hat{M}$ in the semiclassical limit. Coarse-graining was there interpreted
as an average over a set of quantum maps close to identity.
The phase space can be an arbitrary (quantizable) symplectic manifold, and the
classical and quantum averages are generated by a finite set of Hamiltonians
$\{H_j\}_{j=1,\ldots,f}$; $K_\eps$ is a smooth probability kernel in $f$ variables with 
compact support of scale $\eps$ around the origin. The classical and quantum coarse-graining
operators are defined as:
\begin{equation}\label{DZirn}
\begin{split}
D^{\{ H_j\} }_\eps\rho (x)
&=\int _{\R }d^{f}t\, K_{\eps }(t)\;\rho \bigl( \varphi^{-1}_{\sum_j t_j H_j}(x)\bigr) \\
\hat{D}^{\{H_j\}}_\eps\hat\rho
&=\int_{\R^f}d^f t\, K_\eps(t)
\ad \left( \e^{-\i \sum_j\hat{H}_j t_j/\hbar}\right)\rho.
\end{split}
\end{equation}
This scheme is more
general than what we have done on the torus: one does not need any group action
on the phase space, but only a sufficient number of Hamiltonians.
We recover our previous definition on the torus if we take for ``Hamiltonians''
the multivalued functions $H_1=p\bmod 1$, $H_2=-q\bmod 1$ (these
functions are not well-defined on $\t2 $, but their flows are). 
The qualitative spectral
features of the classical coarse-grained propagator $D^{\{H_j\}}_\eps$
for small $\eps $ should not depend on the selected family of Hamiltonians
$\{H_j\} $, as long as the second-order 
operator $\sum_{j=1}^f(\nabla_{H_j})^2$ is elliptic (in the above case
on the torus, this is the Laplace operator 
$\frac{\partial^2}{\partial q^2}+\frac{\partial^2}{\partial p^2}$, which is
indeed elliptic) \cite{kifer}. 

In the framework of quantum mechanics on the 2-sphere, 
a different type of dissipative superoperator $\hat\P_\tau=\hat D_\tau\circ \ad(\hat M)$ was
considered in \cite{braun}, where $\hat D_\tau$ is a quantum dissipation operator 
obtained by integrating a quantum \emph{master equation} during the time $\tau$. 
This dissipation operator first appeared in the study of superradiance
in atomic physics \cite{superradiance}.
As a main difference with our noisy propagator $\hat\P_\eps$, 
the (unique) invariant eigenmode of $\hat\P_\tau$ is different from $\hat\rho_0$. 
This corresponds to the fact that the corresponding 
classical propagator $\P_\tau$ (obtained from $\hat\P_\tau$ by taking the semiclassical
limit) does not leave invariant the uniform 
density $\rho_0$, but rather a more singular measure, 
which may supported either on a discrete set
of points, or even on a more complicated ``strange attractor''.


\subsection{Quantum coarse-grained propagator}

We will now study the quantum analogue of $\P_\eps$, that is
the coarse-grained quantum propagator $\hat\P_\eps=\hat D_\eps\circ\hat\P_M$.
Similarly as the classical propagator, $\hat\P_\eps$ maps
a Hermitian density to a Hermitian density; as a result, its spectrum is 
symmetric w.r.to the real axis. Like $\hat D_\eps$, $\hat\P_\eps$ is a Kraus superoperator, and
therefore realizes a quantum Markov process: the quantum density first evolves
through the deterministic map $\ad(\hat M)$, then performs a
random quantum jump through $\ad(\hat{T}_{V/N})$, with a probability
$\propto K_{\eps }(V/N)$. 

Linear combinations of quantum translations were
considered in \cite{paz-saraceno} as models for decoherence
on the quantum torus. The authors studied the evolution of densities through
an operator similar with $\hat\P_\eps$, by computing the time evolution
of the `purity' $\tr(\hat\rho^2)$. They took for $\hat M$
the quantized Baker's map, which is fully chaotic, yet discontinuous on $\t2$.
More recently, similar computations were performed for smooth nonlinear perturbations
of cat maps \cite{saraceno}. 

The author of \cite{braun}
uses Gutzwiller-type trace formulas to estimate the traces of powers of the 
dissipative quantum propagator $\tr\big(\hat\P_\tau^n\big)$, in
the regime of $n$ fixed and $\hbar\to\infty$; from them he
shows that each trace converges to the corresponding trace of the
classical dissipative propagator $\tr\big(\P_\tau^n\big)$ (this trace being given
by a sum over periodic orbits as well).

In the following I will not use any trace formula, but more basic 
semiclassical and operator-theoretic techniques to 
compare quantum and classical coarse-grained propagators. 
$\hat\P_\eps$ is indeed spectrally similar with its
classical counterpart $\P_{\eps}$. It conserves the trace:
$\hat\P_\eps\hat\rho_0=\hat\rho_0$, but in contrast with its noiseless version
$\hat\P$, it has for unique invariant density $\hat\rho_0$,
all other eigenvalues (in number $N^2-1$) being inside the unit disk.
As a non-classical property, $\hat\P_\eps$ destroys purity: the image
of a pure state $\hat\rho =|\psi\rangle \langle\psi |$ is not a pure
state.

The following theorem, which is the central result of this paper, 
relates more precisely the spectra of 
$\hat\P_\eps$ and $\P_\eps$ in the semiclassical limit: it states the
``semiclassical spectral stability'' of coarse-grained propagators.

\begin{theo}
\label{th: semiclassical spectral stability}For any smooth map $M$ on
the torus and any fixed $\eps >0$, the spectrum of the quantum coarse-grained
propagator $\hat\P_\eps=\hat D_\eps\circ\hat\P_M$ on $L^2_N$
converges in the semiclassical limit $N\to \infty$ to the spectrum
of the classical coarse-grained propagator $\P_\eps=D_\eps\circ\P_M$
on $L^2(\t2)$. For any $r>0$, the convergence is uniform
in the annulus $R_r=\{r\leq |\lambda| \leq 1\}$.
\end{theo}
To compare the classical and quantum propagators, we use the isometry between
the subspace $\I_N$ of $L^2(\t2)$ and $L^2_N$, induced
by the Weyl quantization $Op_N$ and its inverse $W^P$ (see section
\ref{s:weyl quantization} for notations). $\hat\P_\eps$ is then isometric to 
$\sigma_N(\hat\P_\eps)\defi W^P\circ\hat\P_\eps\circ Op_N\circ\Pi_{\I_N}$
($\Pi_{\I_N}$
projects orthogonally $L^2(\t2)$ onto $\I_N$). We therefore need to
compare the operators $\sigma_N(\hat\P_\eps)$ and $\P_\eps$ on $L^2(\t2)$. 
The crucial semiclassical estimate is the following lemma.

\begin{lemm}
\label{lem:operator convergence}
The finite-rank operators $\sigma_N(\hat\P_\eps)$
converge to $\P_\eps$ in the operator norm on $L^2(\t2)$,
in the limit $N\to \infty $.
\end{lemm}
\emph{Proof:}
The key semiclassical ingredient is Egorov's property \eqref{e:Egorov}. 
From the norm inequality 
$$
\Vert \hat\rho\Vert^2_{HS}=\frac{1}{N}\tr (\hat{\rho }^{\dagger }\hat{\rho })
\leq \Vert \hat\rho\Vert^2_{\mathcal{L}(\hn)},
$$
the convergence in (\ref{e:Egorov}) also holds for the Hilbert-Schmidt norm,
that is on the space $L^2_N$. Using the symbol map $W^P$ and its inverse $Op_N$ on $\I_N$,
we will convert operators on $L^2_N$ into operators on $L^2(\t2)$. 
We notice that for any $k\in\Z^2$, $\rho_k\in \I_N$ for large enough $N$. The evolved
density $\P_M\rho_k$ is in general not in $\I_N$, but it is smooth since $M$ is so. 
Any smooth density $\rho$ is asymptotically equal to its projection on $\I_N$, so that 
$$
\forall \rho\in C^\infty(\t2),\quad\lVert W^P\circ Op_N(\rho)-\rho\rVert_{L^2(\t2)}
\stackrel{N\to\infty}{\longrightarrow} 0.
$$
Using this fact, Egorov's property can be recast into:
$$
\forall k\in \Z^{2},\quad 
\Vert \sigma_N(\hat{\P}_M)\rho_k - \P_M\rho_k\Vert_{L^2(\t2)}
\stackrel{N\to\infty}{\longrightarrow} 0.
$$
This means that the sequence of operators $\sigma_N(\hat{\P}_M)$
semiclassically converges to $\P_M$ in the \emph{strong topology} on $\mathcal{B}(L^2)$
(the bounded operators on $L^2(\t2)$).
Now, a standard lemma in functional analysis \cite[Lemma 2.8]{toeplitz} states
that if a sequence $A_n$ of bounded operators on some Banach space converge
strongly to the operator $A$, then for any compact operator $K$, $KA_n$
converges to $KA$ in operator norm. Since $D_{\eps}$ is compact,
this implies that $D_\eps\circ \sigma_N(\hat\P_M)$ converges
to $\P_{\eps,M}$ in the operator norm. A simple comparison of the eigenvalues
shows that $\Vert\sigma_N(\hat D_\eps)-D_\eps\Vert_{\mathcal{B}(L^2)}\to 0$
as $N\to \infty $, which achieves to prove the lemma.
\hfill$\square$

\medskip

\emph{End of proof of the theorem:}
From lemma \ref{lem:operator convergence}, one applies standard methods 
to show that the spectrum of $\sigma_N(\hat\P_\eps)$
converges to that of $\P_\eps$, as was done in Section \ref{s:anosov}.
Namely, for any $\lambda\neq 0$ and any small disk $B(\lambda,\delta)$ around it, 
the spectral projectors for $\sigma_N(\hat\P_\eps)$ and 
$\P_\eps$ in that disk satisfy
$\Vert \Pi^{(N)}_{B(\lambda,\delta )}-\Pi_{B(\lambda,\delta)}\Vert \to 0$ as $N\to\infty$.
For small enough $\delta$, the projector $\Pi_{B(\lambda,\delta)}$ is 
independent of $\delta$ and of finite rank, equal to the multiplicity of $\lambda$ in
the spectrum of $\P_\eps$.
The above convergence implies that 
$\Pi^{(N)}_{B(\lambda,\delta )}$ has the same rank for $N$ large enough,
and that the corresponding eigenspaces of $\P_\eps$ and $\sigma_N(\hat\P_\eps)$ 
are close to each other. 
Finally, the spectrum of $\hat\P_\eps$ is identical with that of 
$\sigma_N(\hat\P_\eps)$.
\hfill$\square$

This spectral stability was noticed numerically in \cite{manderfeld} for the kicked
rotator on the 2-sphere. In their case, the coarse-graining consists in a sharp
truncation of the  Fourier expansion of the Husimi functions of the quantum densities, 
but the same
arguments as above can be applied to show the spectral stability of the
coarse-grained propagator in the semiclassical limit.


\section{On the non-commutativity of the semiclassical vs. small-noise limits
\label{s:non-commute}}

In the previous section we have described the semiclassical limit of the quantum 
coarse-grained propagator $\hat\P_\eps$
for a \emph{fixed} coarse-graining width $\eps>0$. On the other hand, 
Section~\ref{s:spectrum classical} was dealing with the small-noise limit
$\eps \to 0$ for the classical propagator $\P_\eps$. 

These two limits do not commute. 
Indeed, for fixed $N\in\N$ the coarse-graining operator $\hat{D}_\eps$ is
a finite matrix, the $N^2$ eigenvalues of which
converge to unity as $\eps\to 0$ (see Eq.~(\ref{e:d_eps,k})). 
Therefore, in this
limit $\Vert\hat{D}_\eps -\textrm{Id}_{L^2_N}\Vert \to 0$.
For $K(x)$ of compact support, we even have $\hat{D}_\eps=\textrm{Id}_{L^2_N}$ as
soon as the rescaled support $(\eps N)Supp(K)$ has no intersection with $\Z_*^2$.
This shows that for $N$ fixed and $\eps$ decreasing to zero, $\hat\P_\eps$
is close to unitary on $L^2_N$, uniformly with respect to
the map $M$:
\begin{equation}
\label{e:quasi-unitary}
\forall \hat{\rho }\in L^{2}_{N}\ \textrm{ s.t.}\  
\| \hat\rho\|_{_{HS}}=1,\quad (1-\|\hat{D}_\eps -\textrm{Id}_{L^2_N}\|)
\leq \|\hat\P_\eps\hat\rho\|_{_{HS}}\leq 1.
\end{equation}
On the opposite, there should exist a regime where $N\to \infty $ (semiclassical)
and simultaneously $\eps=\eps(N)\to 0$ (vanishing noise) slowly enough, 
such that the eigenvalues
of $\hat\P_{\eps (N)}$ stay close to the eigenvalues of $\P_{\eps (N)}$,
and therefore behave differently according to the classical properties of $M$.
For an Anosov map, the ``outer'' eigenvalues (say, in some annulus $R_r$) will
converge to the Ruelle resonances, while for an integrable map they will form dense strings 
touching the unit circle. Eq.~(\ref{e:quasi-unitary}) shows that a 
\emph{necessary} condition
for $\hat\P_\eps$ to possess eigenvalues close to the origin (like
$\P_\eps$) is that the coarse-graining operator $\hat{D}_\eps$ itself
has small eigenvalues. The smallest eigenvalues of $\hat{D}_\eps$ correspond to the
largest wave vectors in $\Z^2_N$, namely $|k|\sim N/2$. 
From the explicit expression (\ref{e:d_eps,k}), this implies
the condition \begin{equation}
\label{e:necessary condition}
N\eps (N)>>1.
\end{equation}
This condition is quite obvious: it means that the scale of coarse-graining $\eps(N)$
must be larger than the ``quantum scale'' $\frac{1}{N}=2\pi \hbar $, that is 
the distance between two nearby position states $|q_j\rangle_N$. 
One may wonder if this condition is sufficient, or
if $\eps(N)$ should decrease slowlier to get the desired convergence. We have
so far no definite answer to this question for a general map.

In the next two subsections, we will compare the spectra of $\hat\P_\eps$
and $\P_\eps$ for the various maps treated classically in sections
\ref{s:cat classical} and \ref{s:integrable classical}. We know
no nonlinear Anosov map for which Ruelle resonances can be computed analytically, so
for this case  we rely on numerical studies for the perturbed
cat map \cite{agam-blum}.

We plot some numerically computed spectra, always using Gaussian noise
and selecting two different $\hbar$-dependences
for the noise width $\eps(N)$. We consider either a ``slow decrease''
$\eps(N)=N^{-1/2}$, for which the convergence to
classical eigenvalues is checked even for relatively small values of $N$
(the largest value of $N$ we considered is $N=40$). 
To test the finer condition \eqref{e:finer condition},
we also consider a ``fast decrease'' of the noise width
$\eps(N)=\frac{\log(N)}{N}$, the convergence to classical eigenvalues
being then harder to verify numerically. 

\begin{figure}[htbp]
\centerline{\rotatebox{-90}{\includegraphics[width=8cm]{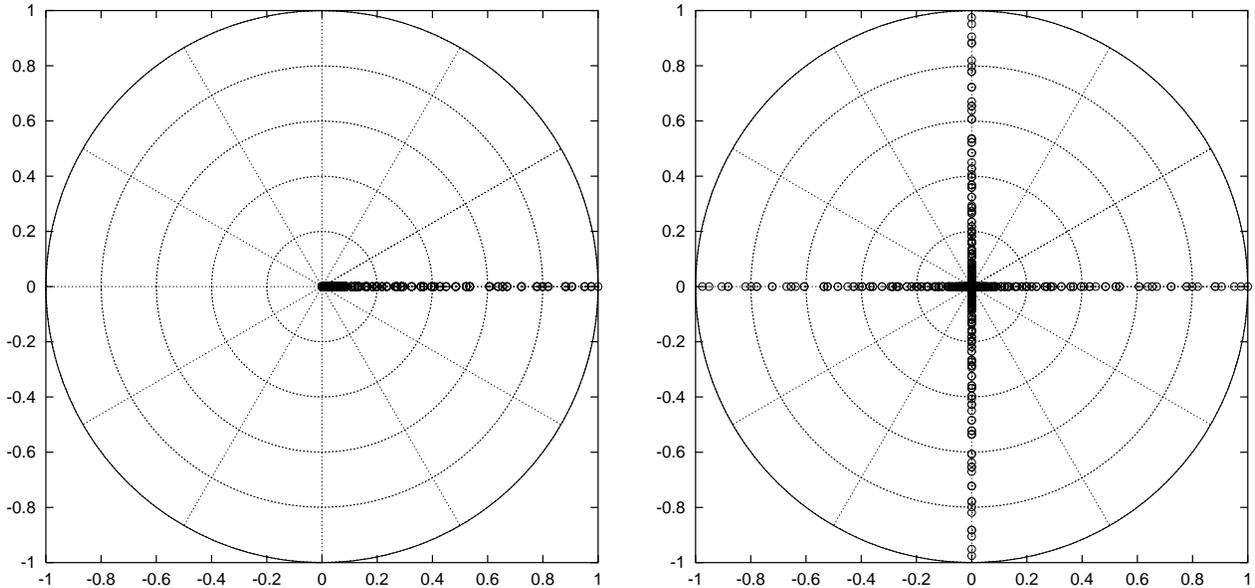}}}
\caption{\small\label{fig:dcas-four-140} Spectrum of the quantum coarse-graining operator 
$\hat{D}_\eps$ for
the Gaussian noise (left); Spectrum of the coarse-grained $\pi/2$-rotation $\hat{\P}_{\eps,J}$ 
(right). Parameters are $N=40$, $\eps=N^{-1/2}$ (small circles). The large concentric
circles are only shown for clarity.}
\end{figure}

We start by plotting on Fig.~\ref{fig:dcas-four-140} (left) the spectrum of the quantum
coarse-graining operator $\hat{D}_\eps$.


\subsection{Quantum linear automorphisms}
In this section, we will only consider linear maps $A$ satisfying the 
``checkerboard condition'' given in Section \ref{s:quantize maps}, necessary for their
quantization on $\hn$.
Then, the quantized linear automorphism $\hat A$ satisfies the 
\emph{exact} Egorov property \cite{hannay}:
$$
\forall k\in \Z^2,\quad \hat{A}\,\hat{T}_{k/N}\hat{A}^{-1}
=\hat{T}_{Ak/N}\Longleftrightarrow \hat\P_A\, Op_N(\rho_k)
=Op_N(\P_A \rho_k)=Op_N(\rho_{Ak}).
$$
The quantum propagator therefore acts as a permutation on the quantum translations,
like the classical propagator on plane waves (cf. Eq.~\eqref{e:permutation}). 
In a first step, we treat any linear map, regardless of the nature of its dynamics.

The main difference between the
quantum and classical frameworks comes from the fact that Weyl quantization $Op_N$
is not one-to-one: $Op_N(\rho_{k+Nk'})\propto Op_N(\rho_k)$ for any $k'\in\Z^2$.
As a result, each orbit $\O(k)=\{ A^t k,\, t\in \Z \}$
has to be taken modulo $\Z_N^2$ in the quantum case, yielding a
``quantum orbit'' $\O_N(k)$ which is necessarily \emph{finite}. 
Through a rescaling of $\frac{1}{N}$,
$\O_N(k)$ is identified with a periodic orbit of the map $A$ on the torus,
situated on the ``quantum lattice'' $(\frac{1}{N}\Z)^2$.
The period $T_{N,k}$ of this orbit is the smallest $t>0$ such that
$A^t k\equiv k\bmod \Z_N^2$. To compute the spectrum of $\P_A$ and $\P_{\eps,A}$ we
need to analyze these quantum orbits.

The quantum
orbits form a partition of $\Z_N^2$, therefore a partition of the
basis $\{\hat{T}_{k/N},\, k\in \Z_N^2\} $ of $L^2_N$.
To each $\O_N(k)$ corresponds the $T_{N,k}$-dimensional subspace
$Span\O_N(k)$, invariant through both $\hat\P_A$ and $\hat{D}_\eps$.
The quantum propagator $\hat\P_A$ satisfies (we 
abbreviate $T_{N,k}$ with $T$)
\begin{equation}
\label{e:quantum period}
\hat\P_A^{T}Op_N(\rho_k)=Op_N(\rho_{A^{T}k})=Op_N(\rho_{k+NV})
=(-1)^\gamma\: Op_N(\rho_k)
\end{equation}
for some $V\in \Z^2$, and $\gamma =k\wedge V+N V_1 V_2$. This
implies that the eigenvalues of $\hat\P_A$ on $Span\O_N(k)$
are the phases 
$\{ \exp \bigl( \frac{2\i \pi }{T}(r+\gamma /2)\bigr),\: r=0,\ldots ,T-1\}$.
By switching on the noise, the equation (\ref{e:quantum period}) is modified
by inserting the action of $\hat{D}_\eps$ on the successive modes $\hat{T}_{A^{t}k/N}$.
As a result, 
$$
\hat\P_{\eps,A}^{T}\, \hat{T}_{k/N}=(-1)^{\gamma }\, d^{(N)}_{\eps ,\O (k)}\: \hat{T}_{k/N},
$$
 with 
$$
d^{(N)}_{\eps ,\O (k)}\defi \prod _{t=0}^{T-1}d^{(N)}_{\eps ,A^{t}k}.
$$
The spectrum of $\hat\P_\eps$ on $Span\O_N(k)$ thus consists
in $T$ regularly spaced points on the circle of radius 
$\vert d^{(N)}_{\eps ,\O (k)}\vert ^{1/T}$. 
To estimate this radius, we take its logarithm
\begin{equation}
\label{e:d_O(k)}
\frac{1}{T}\log |d^{(N)}_{\eps,\O(k)}|
=\frac{1}{T}\sum _{k'\in \O_N(k)}\log |d^{(N)}_{\eps ,k'}|.
\end{equation}

According to Eq.~(\ref{e:d_eps,k}), this quantity is the average over the periodic
orbit $\frac{1}{N}\O _{N}(k)$ of the function 
$f_{K}(\eps N,\zeta )\defi 
\log \left| \frac{\theta_{K}(\eps N,\zeta )}{\theta _{K}(\eps N,0)}\right| $.
This function is smooth in $\zeta$ except at possible logarithmic singularities
if $\theta_{K}$ vanishes. In the Gaussian case $G(x)=\e ^{-\pi |x|^{2}}$,
$f_{G}$ has no singularities and admits for $\eps N>>1$
the asymptotic behaviour 
$f_{G}(\eps N,\zeta )\sim -\pi(\eps N\zeta )^{2}$
in the square $\{ |\zeta _{1}|\leq \frac{1}{2},\, |\zeta _{2}|\leq \frac{1}{2}\}$.

We have obtained an explicit relationship between the spectrum of $\hat\P_{\eps ,A}$
and the structure of periodic orbits of $A$ on the quantum lattice. The latter
drastically differs between chaotic and integrable automorphisms, which leads to 
qualitatively different quantum spectra. Below,
we sketch the description of these quantum orbits, respectively for the elliptic,
parabolic and hyperbolic maps. We use the notations and results of
Sections \ref{s:cat classical}, \ref{s:non-mixing automorphism}.


\subsubsection{Elliptic transformation \protect$J\protect$\label{elliptic}}

The quantization of the $\pi /2$-rotation $J$ yields the finite Fourier transform $\hat J$.
The classical orbits
$\O(k)$ are of period $4$. In general, the successive $J^{j}k$
are not congruent $\bmod N$, so that the quantum orbit $\O_N(k)$
also has period $4$. The only exception occurs for $N$ even, $k=(\frac{N}{2},\frac{N}{2})$
(period $1$) or $k=(0,\frac{N}{2})$ (period $2$). Assuming that the
coarse-graining kernel has the symmetry $\tilde{K}(\xi)=\tilde{K}(J\xi)$,
the eigenvalues of $\hat\P_{\eps,J}$ on a 4-dimensional $Span\O_N(k)$
are $\{ \i ^{l}d^{(N)}_{\eps ,k}\, ,\, l=0,\ldots ,3\}$. Taking
all quantum orbits into account, the spectrum forms $4$ strings as in the
classical case (see Fig.~\ref{fig:dcas-four-140}, right). 
For a Gaussian noise, Eq.~(\ref{e:quantum Gaussian case})
shows that these eigenvalues are exponentially close to the corresponding classical
eigenvalues.


\subsubsection{Parabolic linear shear $S$ \label{s:quantum shear}}
We can quantize on $\hn$ the parabolic shear $S$ described in 
Section~\ref{s:non-mixing automorphism} if the integer $s$ is \emph{even}.
Quantizing the space of $\P_S$-invariants $V_{inv}=Span\{\rho_{(l,0)},\, l\in \Z\}$
leads to the $\hat\P_S$-invariants 
$V_{inv,N}=Span\{\hat{T}_{(\frac{l}{N},0)},\, l\in \Z_N\}$.
Each $\hat{T}_{(\frac{l}{N},0)}$ is eigenstate of $\hat\P_\eps$
with eigenvalue $d^{(N)}_{\eps,(0,l)}$, which is close to the classical eigenvalue
$\tilde{K}(\eps(0,l))$ for $\eps N>>1$; these
eigenvalues then form a string connecting unity to the origin.

\begin{figure}[tbp]
\centerline{\rotatebox{-90}{\includegraphics[width=8cm]{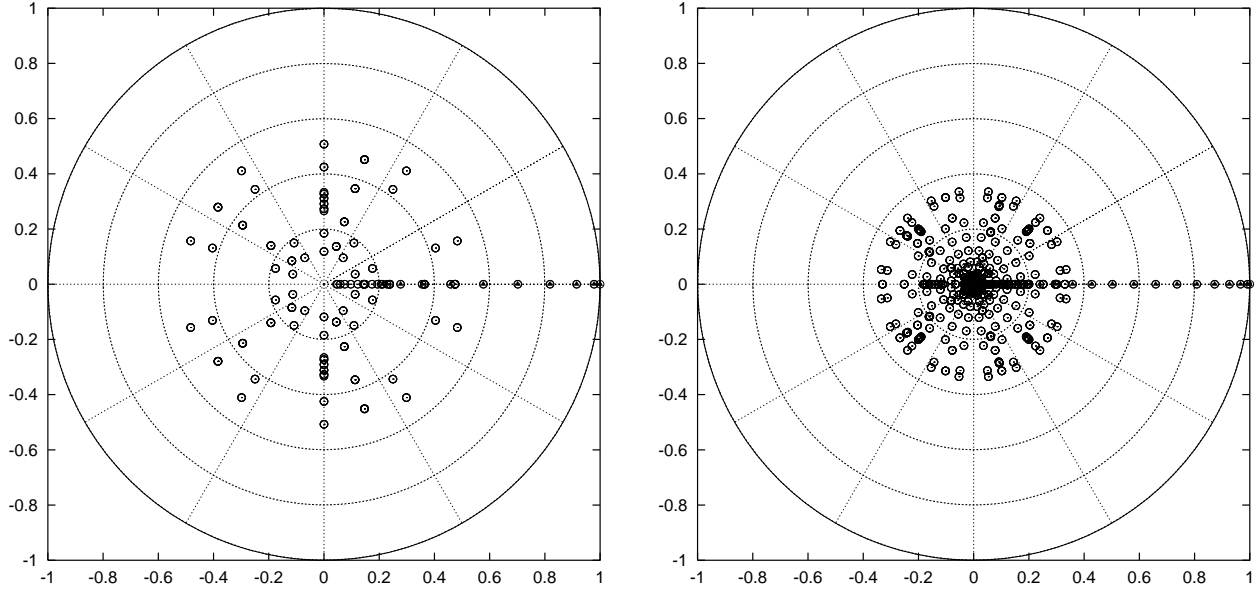}}}
\caption{\label{fig:slinlog20-40}\small Spectrum of $\hat{\P}_{\eps,S_2}$ for
the linear shear $S_2$ (circles) and values $\{d^{(N)}_{\eps,(j,0)}\}_{j=0}^N$ (triangles). 
Parameters are $N=20$ (left), $N=40$ (right) and in both cases $\eps=\log(N)/N$.}
\end{figure}

Now we study the spectrum of $\hat\P_\eps$ on the orthogonal of the
invariant space, $V_{inv,N}^\perp$. For $k=(k_1,k_2)$ with
$k_2\not \equiv 0\bmod N$, the infinite orbit $\O(k)=\{k+(jsk_2,0),\,j\in \Z \}$
projects modulo $N$ onto a finite orbit $\O_N(k)$ whose period depends on the ``step''
$g\defi\gcd (N,sk_2)$: 
$\O_N(k)=\{ (k_1+jg\bmod N,k_2)\: \vert j=0,\ldots,\frac{N}{g}-1\}$.
For a fixed $k_2\in \Z_{*}$, the step $g$ stays bounded in the limit of large
$N$, so that the sum (\ref{e:d_O(k)}) behaves as the integral 
$\int_0^1 dt\, f_K\left( \eps N,(t,\frac{k_2}{N})\right)$;
for the Gaussian noise, the latter yields $-\pi\eps^2(\frac{N^2}{12}+k_2^2)+o(1)$.
For a general value of $k_2\in\Z_N\setminus 0$, 
the step $g$ may be of order $N$, in which case
the sum (\ref{e:d_O(k)}) is not well-approximated
by an integral; still, one can (for Gaussian noise) prove the uniform upper bound: 
$$
\forall k\in\Z_N^2\setminus 0,\quad\frac{1}{T_{N,k}}\log |d^{(N)}_{\eps,\O (k)}|\leq -C(\eps N)^{2}
$$
with $C=\pi \min \{ 1/s^{2},1/16\}$. The eigenvalues
of $\hat\P_\eps$ on $V_{inv,N}^\perp$ are therefore situated on circles
of radii $\leq\e^{-C(\eps N)^2}$, so they
uniformly converge to zero as $\eps N\to\infty$ (we remind that the spectrum of
$\P_{\eps,S}$ restricted to $V_{inv}^\perp$ reduces to $\{0\}$). 

In Fig~\ref{fig:slinlog20-40} we show the spectra of $\hat{P}_{\eps,S_2}$ for the 
linear shear $S_2=\begin{pmatrix}1&2\\0&1\end{pmatrix}$ and a ``fast decreasing noise'', 
together with the eigenvalues $\{d^{(N)}_{\eps,(j,0)}\}_{j=0}^N$. 
The `non-invariant spectrum' converges slowly to the origin, while the 
`invariant spectrum' becomes dense on the interval $[0,1]$. Had we chosen a larger
noise width, the non-invariant spectrum would be contained in a smaller disk for the same 
values of $N$.


\subsubsection{Hyperbolic transformations}
The space of invariant densities for a hyperbolic automorphism $A$ reduces to 
$V_{inv}=\C\rho_0$.
Its orthogonal $V_{inv}^{\perp}=L^2_0$ is quantized into the space $L^2_{0,N}$
of traceless densities. We remind (cf. Section~\ref{s:cat classical}) 
that the spectrum of $\P_{\eps,A}$ on this
subspace reduces to the $\{0\}$, for any $\eps>0$.

The periodic orbits of a hyperbolic automorphism $A$ 
were thoroughly studied in \cite{PV}; the authors classified
the orbits according to the $1/N$ sublattice they belong to. Yet, they gave no
equidistribution estimate on long periodic orbits.
Our aim is to estimate the RHS of Eq.~(\ref{e:d_O(k)}) for
all quantum orbits, at least for large enough $N$ (we will restrict ourselves
to the Gaussian coarse-graining). For any Anosov map, the long periodic orbits
equidistribute in the statistical sense (averaging over all orbits
of a given period) in the limit of long periods \cite{PP}. 
For a certain class of hyperbolic automorphisms, 
semiclassical equidistribution of \emph{all} quantum orbits was obtained 
for an infinite subsequence of values of $N$
\cite{DEGI}. Equidistribution morally implies that the sum (\ref{e:d_O(k)}) 
behaves as the integral 
$\int_{\t2 }dx\, f_{G}(\eps N,x)\approx \frac{\pi(\eps N)^{2}}{6}$.
One can indeed show that for $N$ in this subsequence,
any eigenvalue $\lambda$ of $\hat\P_{\eps,A}$ on the subspace $L^2_{0,N}$ satisfies
$$|\lambda|\leq \exp \left\{ -\frac{\pi (\eps N)^{2}}{6}+\O (\eps ^{2}N^{3/2})\right\}.$$ 

\begin{figure}[tbp]
\centerline{\rotatebox{-90}{\includegraphics[width=8cm]{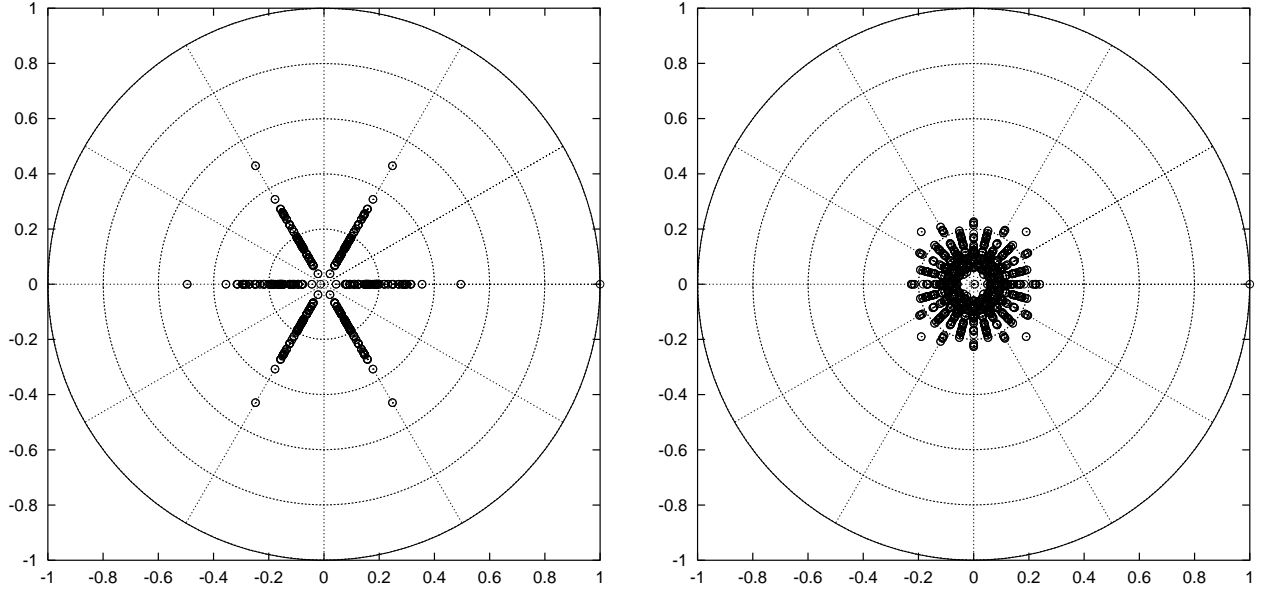}}}
\caption{\small\label{fig:catlog30-40} Spectrum of $\hat{\P}_{\eps,A_0}$ for
the cat map $A_0$. 
Parameters are $N=30$ (left), $N=40$ (right) and in both cases $\eps=\log(N)/N$. $N=30$
corresponds to a ``short quantum period'' $T(30)=6$, which is clearly visible on
the shape of the spectrum.}
\end{figure}

There also exist arbitrary large values of $N$ for which the period $T(N)$
of $A$ modulo $N$ is as small as $\frac{2\log N}{\lambda }+\O (1)$,
where $\lambda>0$ is the Lyapounov exponent of $A$ \cite{PV}.
All quantum orbits $\O_N(k)$ have periods dividing $T(N)$. Starting from
$k_0=(0,1)$, the linear dynamics shows that the point $\frac{A^{t}k_0}{N}$
remains close to the origin (that is, at a distance $<<1$) along the unstable
direction until the time $\approx \frac{\log N}{\lambda}$, when it reaches
the boundary of the square $\{|\zeta_1|\leq 1/2,\,|\zeta_2|\leq 1/2\}$; 
it is then straight away `captured'
by the stable manifold, which brings is back to the origin during the remaining
$\approx\frac{\log N}{\lambda}$ steps. This orbit thus achieves an ``optimally
short'' homoclinic excursion from the unstable origin, and is far from being
equidistributed. The average of $f_{G}$ along this orbit is of order 
$-\frac{C(\eps N)^{2}}{\log N}$,
so the eigenvalues of $\hat\P_{\eps,A}$ on $Span\O_N(k_0)$
have a radius 
$\approx\exp \bigl( -\frac{C(\eps N)^{2}}{\log N}\bigr)$.
These eigenvalues will therefore semiclassically converge to zero under the condition
\begin{equation}\label{e:finer condition}
\frac{\eps N}{\sqrt{\log N}}\to \infty.
\end{equation}
We believe that these particular values of $N$ represent the ``worst case'' 
as far as equidistribution of
long orbits is concerned, and that condition \eqref{e:finer condition} suffices for the
spectra of $\hat\P_{\eps,A|L^2_{0,N}}$ to semiclassically converge to zero 
for the full sequence $N\in\N$. In Fig.~\ref{fig:catlog30-40} we show the spectrum of 
$\hat\P_{\eps,A_0}$ for the quantization of
the cat map $A_0=\begin{pmatrix}2&1\\3&2\end{pmatrix}$, with the fast 
decreasing noise width.

\paragraph{Remark:}
For both the parabolic and the hyperbolic automorphisms, the spectrum of $\hat\P_\eps$ on 
$V_{inv,N}^\perp$ reduces to $\{0\}$ if the coarse-graining consists in a sharp
truncation in Fourier space, and $\eps^{-1}$ grows as $\eps^{-1}\approx c N$ with  
$c$ a finite but small constant (depending on the classical map). Any nontrivial quantum 
orbit $\O_N(k_0)$ then contains an element $k$ s.t. $|k|>\eps^{-1}$,
such that the corresponding mode $\hat T_{k/N}$ is killed by $\hat D_\eps$.


\subsubsection{Quantum translations}

As we mentioned in Section \ref{s:quantize maps}, any classical
translation $T_v$ with $v\in\t2$ can be quantized on $\hn $
by the quantum translation $\hat{T}_{v^{(N)}}$, where $v^{(N)}$ is
the ``closest'' point to $v$ on the quantum lattice. 
This quantization was shown \cite{marklof}
to satisfy Egorov's property (\ref{e:Egorov}).

From Eq.~\eqref{e:translation group}, the quantum propagator 
$\hat\P_{v^{(N)}}=\ad (\hat{T}_{v^{(N)}})$
admits any quantum translation $\hat{T}_{k/N}=Op_{N}(\rho _{k})$ as an eigenstate,
with eigenvalue $\e^{-2\i\pi v^{(N)}\wedge k}$. All eigenvalues
are $N$-th roots or unity, and are at least $N$-degenerate; in case the
classical translation $T_v$ is ergodic, these degeneracies contrast with the
nondegenerate (but dense) spectrum of $\P_v$. On the opposite, for a rational translation
(the classical eigenvalues form a finite set), the quantum eigenvalues may take values in 
all $N$-th roots of unity, in case $N$ and $v^{(N)}$ are coprime. The spectra of 
$\P_v$ and $\hat\P_{v^{(N)}}$ may thus be qualitatively very different. 

This difference disappears when one switches on the noise.
Each $\hat{T}_{k/N}$ is also an eigenstate of the coarse-grained propagator 
$\hat\P_{\eps,v}=\hat{D}_{\eps }\hat{\P }_{v^{(N)}}$,
associated to the eigenvalue $\e^{2\i \pi k\wedge v^{(N)}} d^{(N)}_{\eps ,k}$.
In the semiclassical limit, the deviation from the corresponding classical eigenvalue
$\e^{2\i \pi k\wedge v} \tilde{K}(\eps k)$ (cf. Section \ref{s:classical translation})
depends on both the wavevector $k$ and the difference $v-v^{(N)}$.

\begin{figure}[tbp]
\centerline{\rotatebox{-90}{\includegraphics[width=5.5cm]{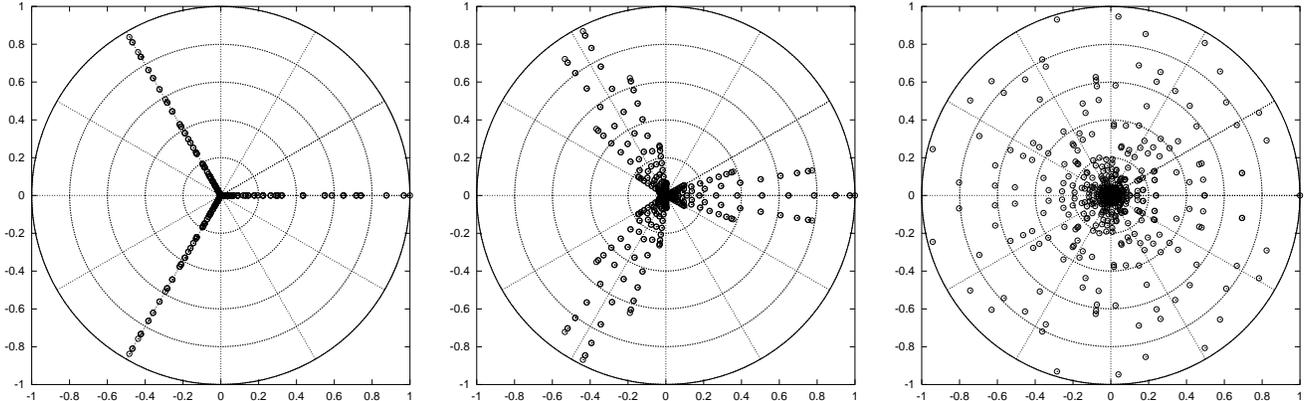}}}
\caption{\small\label{fig:trans30-37-l37} Spectrum of $\hat{\P}_{\eps,v}$ for
the translations $v=(0,1/3)$ (left: $N=30$, center: $N=37$) and $v=(1/\sqrt{2},1/\sqrt{5})$ 
(right: $N=37$). The noise strength is $\eps=N^{-1/2}$.
For the rational translation, the eigenvalues are either exactly on the 
classical axes (for $N$ multiple of $3$), or
semiclassically converge to it. For the irrational one, the eigenvalues become dense in the full disk.}
\end{figure}

To give an example, the rational translation vector $v=(0,\frac{1}{3})$
leads to $3$ eigenvalues $\e^{2\i \pi l/3}$ for the
classical propagator $\P_{v}$,
and $3$ (infinite) strings for the spectrum of $\P_{v,\eps}$. Quantum-mechanically,
if $N$ is a multiple of $3$ one takes $v^{(N)}=v$, and the eigenvalues
of $\hat\P_{v,\eps }$ are exponentially close to the classical ones 
(Fig.~\ref{fig:trans30-37-l37}, left).
In the case $N=3n+1$, the quantum translation vector will be $v^{(N)}=(0,\frac{n}{3n+1})$,
so that the classical and quantum noiseless eigenvalues for
the mode $\rho_k$ deviate of an angle
$2\pi k\wedge (v-v^{(N)})=\frac{2\pi k_1}{3N}$: this deviation can be as large
as $\pi /3$ for wavevectors $|k_1|\approx N/2$. After switching on the 
noise, the classical and quantum eigenvalues
for $k\in\Z_N^2$ can deviate by at most $\O (\frac{1}{\eps N})$, 
the maximal deviations
occurring for wavevectors $|k_1|\simeq \eps^{-1}$, $k_2=\O(1)$ 
(Fig.~\ref{fig:trans30-37-l37}, center).  

\subsection{Examples of quantized nonlinear maps}

In this section we shortly review the spectrum of quantized coarse-grained propagators
for three nonlinear maps: we first treat the nonlinear parabolic shear of section
\ref{s:nonlinear shear classical} and the stroboscopic map
of the Harper Hamiltonian (Section \ref{s: Harper classical}), for which we have
some analytic handle. We then consider a perturbation of the cat map $A_0$ and
compare the numerically computed eigenvalues of $\hat\P_\eps$ to the
spectrum of $\P_\eps$ obtained in \cite{agam-blum}.

\begin{itemize}
\item the nonlinear shear described in Section \ref{s:nonlinear shear classical}
is quantized by taking the product of the quantum linear shear $\hat{S}$
(see Section \ref{s:quantum shear}) with the matrix $\e^{-\i \hat{F}/\hbar }$.
Since $F(p)$ is a function of the impulsion only, its Weyl quantization $\hat{F}$
acts diagonally on the impulsion basis $\{ |p_j\rangle\} $. 
As a result, the perturbation $\ad(\e^{-\i \hat{F}/\hbar }) $
acts trivially on any projector $|p_j\rangle\langle p_j|$, and therefore
on any translation $\hat{T}_{(0,m)/N}$. As a result, the spectrum
of the noisy nonlinear propagator on $V_{inv,N}$ forms the same string as
for the linear shear. 

The action of $\hat\P$ on $V_{inv,N}^\perp$
can be studied as in the classical case: 
the propagator acts inside each subspace
$V_{n,N}=Span\{\hat{T}_{(n,m)/N},\, m\in \Z_N\}$, as a unitary
$N\times N$ Toeplitz matrix $\hat\P^{(n)}$ (instead of a simple permutation). 
In Appendix \ref{a:quantum nonlinear shear}
we study the non-unitary spectrum of  $\hat\P^{(n)}_\eps$ when taking for
$\tilde K(\xi)$ a sharp cutoff: $\hat\P^{(n)}_\eps$ is then the truncation of $\hat\P^{(n)}$ 
to the subspace
$\{|m|\leq \eps^{-1}\}$. We compare this truncated
propagator with the corresponding classical one, and show that if $\eps N>>1$, 
both spectra belong to the same union of 1-dimensional strings contained in
a fixed `small' disk around the origin, the size of which depends on the strength of the
perturbation $F'$. The spectrum of $\hat{\P}$ for the quantized
nonlinear shear $\e^{-\i\hat{F}/\hbar}\hat{S}_2$ with 
$\hat{F}=\frac{0.25}{2\pi}\cos(2\pi\hat{p})$ is shown in Fig.~\ref{fig:snlin-harper40} (left)
for $N=40$ and $\eps=\log(N)/N$, together with the values $\{d^{(N)}_{\eps,(j,0)}\}_{j=0}^N$.

\item The quantization of the stroboscopic map $\varphi_H^1$ for the Harper flow, 
{\it i.e.} the
unitary matrix $\e^{-\i\hat{H}/\hbar}$, leads to the propagator $\hat\P_H$
leaving invariant any density of the type $\hat{H}^n$.
Under the same conditions for the
coarse-graining kernel $\tilde K(\xi)$ as in the classical case
(see Section \ref{s: Harper classical}),
 $\hat\P_{H,\eps}$ may admit for eigenstates $\hat{H}$,
$(\hat{H}^2-\textrm{Id})$ and $\bigl(\hat{H}^3-\frac{7+2\cos(2\pi/N)}{4}\hat{H}\bigr)$, and
the associated eigenvalues $\{d^{(N)}_{\eps,(j,0)}\}_{j=1}^3$ 
converge to the classical eigenvalues 
$\{\tilde{k}(j\eps)\}_{j=1}^3$ if $\eps N\to\infty$. The spectrum of $\hat\P_{H,\eps}$ for
$N=40$ and $\eps=N^{-1/2}$ is shown in Fig.~\ref{fig:snlin-harper40} (right), together
with the values $\{d^{(N)}_{\eps,(j,0)}\}_{j=0}^4$. For the Gaussian kernel used there, only
$\hat H$ is an eigenstate of $\hat D_\eps$, so the subsequent eigenvalues are hybridized by 
the coarse-graining. A ``string'' of eigenvalues along the positive real axis is clearly visible on
the plot.

\begin{figure}[tbp]
\centerline{\rotatebox{-90}{\includegraphics[width=8cm]{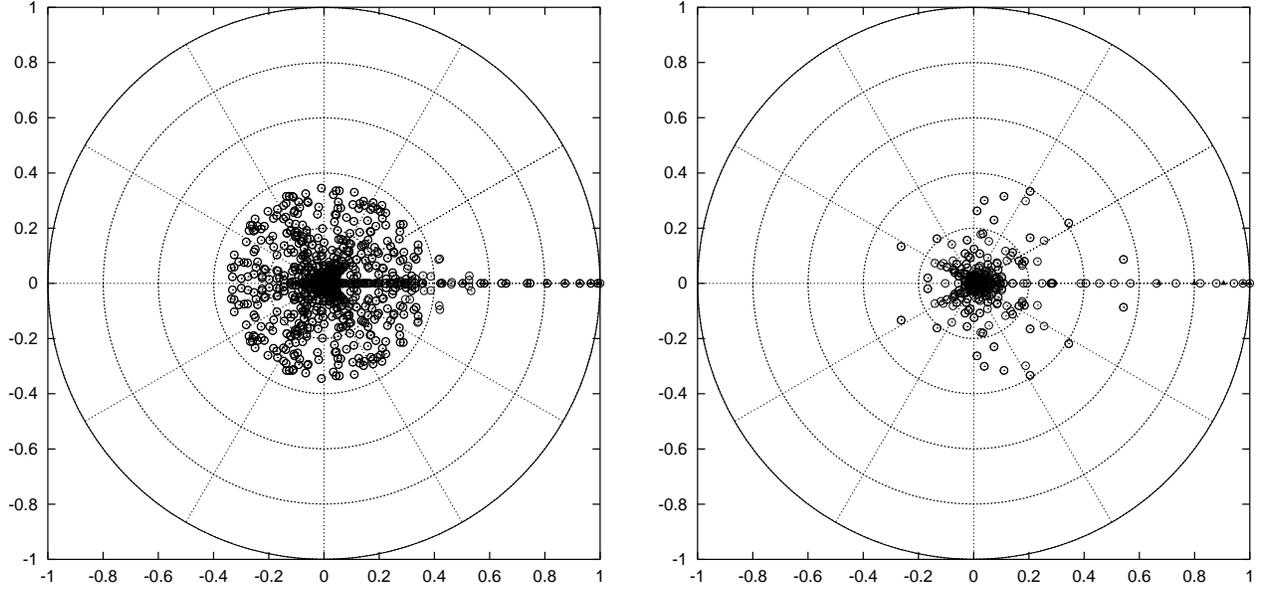}}}
\caption{\small\label{fig:snlin-harper40} Spectrum of $\hat\P_\eps$ for
the nonlinear shear $\e^{-\i\hat F/\hbar}\hat S_2$ (left, $N=40$, $\eps=\log(N)/N$) and the Harper map
$\e^{-\i\hat{H}/\hbar}$ (right, $N=40$, $\eps=N^{-1/2}$). In both cases we also
plotted some values $d^{(N)}_{\eps,(j,0)}$ (black triangles).}
\end{figure}

\item We consider the quantization of the perturbed cat map studied in \cite{agam-blum}, that
is the map $A_{\rm nl}:\binom{q}{p}\mapsto\binom{2q+p}{3q+2p+\kappa(\cos(2\pi q)-\cos(4\pi q))}$. We 
selected the perturbation $\kappa=0.5/2\pi$ in order to compare the quantum spectrum 
with the classical resonance spectrum described in \cite[Fig.~2]{agam-blum}. 
In Figure~\ref{fig:agam40class} we plot the spectrum of $\hat\P_\eps$ for $N=40$
together with the
$7$ `outer' resonances whose values are given in \cite[Table~I]{agam-blum}. In the
`large noise' regime (left), the largest
quantum eigenvalues are indeed close to these resonances, and the rest of the 
spectrum inside a smaller disk. The small-noise regime (right) does not uncover the
classical resonances, however the non-invariant spectrum is already contained in a relatively
small disk around the origin, of radius comparable with the largest resonance.

\end{itemize}

\begin{figure}[tbp]
\centerline{\rotatebox{-90}{\includegraphics[width=8cm]{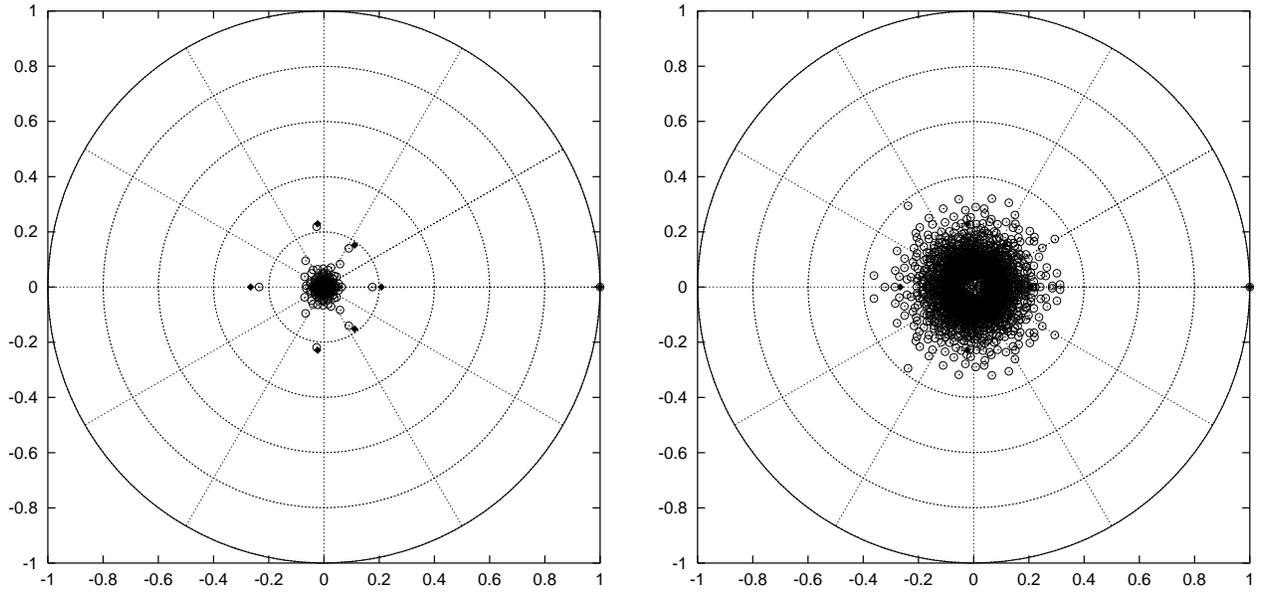}}}
\caption{\small\label{fig:agam40class} Spectrum of $\hat \P_\eps$ for
the quantum perturbed cat map $\hat A_{\rm nl}$ (empty circles)
together with the largest $7$ classical resonances (black diamonds). 
Left: $N=40$, $\eps=N^{-1/2}$. Right: $N=40$, $\eps=\log(N)/N$.}
\end{figure}


\subsection*{Concluding remarks}
In spite of the relatively low values of $N$ used in the numerical plots, 
one can already clearly distinguish two different behaviours of the quantum spectra, depending
on the classical motion, especially for the slow noise decrease $\eps=N^{-1/2}$. For the 
integrable maps, the largest nontrivial eigenvalue
of the noisy quantum propagator $\hat\P_\eps$ is at a distance $\eps^2$ from unity, and
this eigenvalue is the first one of a string of eigenvalues connecting unity with a 
neighbourhood of zero. On the opposite, for the
two chaotic maps we studied (linear and perturbed cat maps), the spectrum already shows a 
finite gap for these values of $N$, even
in the fast-decreasing noise regime $\eps=\log(N)/N$; however, for the relatively low
values of $N$ shown in the plots, 
the gap is governed by
the classical resonances only in the regime of larger noise $\eps=N^{-1/2}$. I conjecture that
the quantum spectrum also converges to the classical spectrum (in any annulus $R_r$)
in the regime $\eps=\log(N)/N$, this  being visible only for much higher
values of $N$.


\section*{General conclusion}
In this paper we have shown the connection between the spectra of coarse-grained quantum and classical 
propagators for maps on the 2-dimensional torus.
I claim that Theorem~\ref{th: semiclassical spectral stability} can be 
straightforwardly extended
to maps on any compact phase space, like the 2-sphere $S^2$ used in \cite{manderfeld,khodas}.
Using our knowledge of the spectrum of the classical noisy map, one infers the presence or the
absence of a \emph{finite gap} between the (trivial) eigenvalue unity on the one hand, the rest of the 
spectrum on the other hand, taking both the semiclassical ($\hbar\to 0$) and small noise 
($\eps\to 0$) limits in a well-defined way.
The presence of this gap in the classical noisy propagator is related with the ergodic
properties of the map. On one extreme, corresponding to integrable maps with infinitely many
invariant densities, there is no gap in the small-noise limit, and the 
spectrum contains a dense `string' of eigenvalues connecting unity with the interior of
the disk: the long-time evolution is therefore governed by the eigenmodes associated with these large
eigenvalues, and is typically of \emph{diffusive type}. On the opposite extreme, for an 
Anosov map (``strongly chaotic''), the spectrum exhibits a finite gap, responsible for
the (classical) exponential mixing. 
The case of maps with less chaotic behaviour is far from being settled. 
The example of irrational translations 
shows that ergodicity does not imply the presence of a gap. Exponential mixing 
was recently 
proven for some non-uniformly or partially hyperbolic maps (see \cite{Baladi} for a review
of recent results), which should (?) imply a gap in the
spectrum of $\P_\eps$. Maps with \emph{subexponential} mixing would
also deserve to be studied from this point of view. The more general 
case of maps with \emph{mixed dynamics}
(where the phase space can be divided between ``regular islands'' and a
``chaotic sea'') was mostly investigated numerically \cite{khodas,manderfeld}; one found the
presence of both ``dense strings'' of eigenvalues and isolated resonances in the spectrum of $\P_\eps$. 
This (physically relevant)
type of dynamics certainly deserves further study.

Like in the classical framework, the presence of a gap in the spectrum of the quantum noisy propagator 
allows to describe the long-time dynamics of densities induced by this propagator:
$\hat\rho(t)= (\hat\P_\eps)^t\hat\rho(0)$. This dynamics is a possible model for a 
dissipative perturbation
of the unitary evolution $\rho\mapsto \hat M^t\rho\hat M^{-t}$, the dissipation being induced by the
interaction with the environment \cite{paz-saraceno}. One measure of the decoherence occurring 
through this evolution is the ``purity'' $\tr(\hat\rho(t))^2$. For an
Anosov map $M$, the long-time evolution
of this quantity will be governed by the largest nontrivial eigenvalues of $\hat\P_\eps$,
which are semiclassically close to the classical resonances. 
This very evolution was recently studied for perturbed
quantum cat maps, providing an estimate of the largest eigenvalues of $\hat\P_\eps$ 
for large values of $N$ \cite{saraceno}. Another way to characterize the time evolution of $\P_\eps$
is through the \emph{dissipation time} 
$t_{diss}=\inf\{t\in\N/\lVert\P_\eps^t\rVert_{\mathcal{B}(L^2_0)}<\e^{-1}\}$. In the case
of linear automorphisms on the $d$-dimensional torus, the small-noise behaviour of $t_{diss}$ 
has been shown to qualitatively depend on the dynamics 
\cite{fannjiang}. The same conclusion should apply to non-linear maps as well.

Our noise operator consists in an average over random translations in
phase space. To diagonalize this operator, we have used the fact that the 
classical (resp. quantum) propagators for translations $\P_v$ (resp. $\hat\P_v$) form a commutative
algebra, and admit as eigenfunctions the Fourier modes. We obtained explicit
expressions for the spectrum of $\P_{\eps,M}$ for the class of linear maps $M$, because
these maps acted simply on these Fourier modes. However, as noticed in Section 
\ref{generalization}, one may be interested in a more general type of diffusion 
operator, like the one defined in Eq.~\eqref{DZirn}. The spectrum of $D^{\{H_{j}\}}_\eps\circ\P_M$ 
for a given map $M$ should be 
qualitatively independent on the family of Hamiltonians $\{H_{j}\}$, as long as the
operator $D^{\{H_{j}\}}_\eps$ has the same characteristics as the coarse-graining $D_\eps$,
namely it leaves almost invariant some ``soft modes'', while killing fast-oscillating
modes. In particular, for
$M$ an Anosov map, the outer spectrum of $D^{\{H_{j}\}}_\eps\circ\P_M$ should converge to 
the set of resonances $\{\lambda_i\}$, and the rest be contained in a smaller disk, in the
limit $\eps\to 0$. For an integrable map, the eigenvalues will not pointwise converge to the ones
of $\hat{D}_\eps\hat{\P}_M$, but they should also accumulate along a `string' touching unity.

Our choice for the superoperator $\hat D_\eps$ was inspired by its classical analogue, as
well as its relevance to modelize the ``quantum noise''.
However, in some quantum systems experiencing (weak) interaction with the environment, 
the effective
``noise'' or ``decoherence'' superoperator may have no obvious 
classical analogue; this seems to be the case for instance in Nuclear Magnetic Resonance 
experiments aiming at
realizing a ``quantum computer'' \cite{tomography}. The quantum Hilbert space is composed of a 
sequence of $n$ spins, so it is isomorphic with the torus Hilbert space $\hn$ for
$N=2^n$ (each spin corresponds to a \emph{binary digit} of the position $q_j$). 
The decoherence operator acting on such a system (for a small number of spins)
has been experimentally measured, it resembles the product of operators acting 
independently on the individual spins, and therefore does not enter in the family of 
diffusion operators described in the present article; 
in particular, such a decoherence operator seems to have 
no obvious classical counterpart.
Conjugating this type of decoherence
operator with a unitary evolution (typically, the finite Fourier transform $\hat J$) 
may lead to a drastically different spectrum from the one shown in Fig.~\ref{fig:dcas-four-140} (right), 
not excluding the presence of a gap. 

\medskip

\textbf{Acknowledgments:} I thank M. Zirnbauer and the Institut f\"ur theoretische
Physik, Universit\"at zu K\"oln, where this work was initiated.
I am grateful to V. Baladi and S. De Bi\`evre for respectively pointing out references \cite{BKL} 
and \cite{fannjiang} to me, and
thank S. Keppeler for his insight on integrable maps.
I had interesting discussions on the whole subject with M. Saraceno and I. Garcia Mata.


\appendix
\section{Appendix}
\subsection{Classical nonlinear shear: spectrum of truncated Toeplitz matrices\label{a:toeplitz}}

This appendix describes the spectrum of the noisy classical propagator for a
nonlinear shear (Section \ref{s:nonlinear shear classical}). In each 
invariant subspace 
$V_{n}=\{\e ^{2\i \pi nq}\, \rho (p)\}$
($n\in \Z^* $), the Perron-Frobenius operator $\P$ acts as
a multiplication of the function $\rho(p)$ on the circle
$\T^1\owns t=\e^{2\i\pi p}$ by: 
\begin{equation}
\label{e:function a_n}
a_{n}(\e ^{2\i \pi p})\defi \e ^{-2\i \pi n(sp+F'(p))}
=\sum _{m\in \Z }\tilde{a}_{n}(m)\e ^{2\i \pi mp}.
\end{equation}
Notice that $a_n(t)=(a_1(t))^n$. In the Fourier basis,
this multiplication has the form of an infinite Laurent matrix $L(a_{n})$, with entries
$L(a_n)_{ij}=\tilde{a}_n(i-j)$. From standard results, the
spectrum of $\P_{|V_n}$ is absolutely continuous, with support the range of the
function $a_n(t)$ (here, the unit circle). 

We consider the coarse-graining consisting in a sharp truncation
in Fourier space: $\tilde K(\xi)=\Theta(\xi_1)\Theta(\xi_2)$. 
The noisy propagator $\P_\eps$ then acts on $V_n$ as
the matrix $L(a_n)$ truncated to the block $\{|m|,|m'|\leq \eps^{-1}\}$, 
or equivalently
as $\{1\leq m,m'\leq 2\eps^{-1}+1\}$. We assume that the smooth function $a_n(t)$
on $\T^1$ can be continued to an analytic (or meromorphic)
function on $\C^*$ (for the linear shear, one has $a_n(z)=z^{-sn}$).
Then, for
any $r>0$ and any $\eps>0$, the spectrum of this truncated matrix is
contained in the \emph{convex hull} of the curve $a_n(r\T^1)$ 
\cite[Prop.~2.17 and Section~5.8]{toeplitz}. For the linear
shear, these curves are
the circles centered at the origin,
so the spectrum reduces to $\{0\}$, as expected. If the perturbation
$F'$ is small, these curves are deformations of circles, some of which remain
in a small neighbourhood of the origin. 
As a concrete example, the perturbation $F(p)=\frac{\alpha}{2\pi}\cos(2\pi p)$
leads to the function $a_{-1}(z)=z^{s}\exp \left\{\pi \alpha (1/z-z)\right\}$.
The curves $a_{-1}(r\T^1)$ satisfy
$$
\forall r>0,\ |z|=r\Longrightarrow |a_{-1}(z)| \leq r^{s}\exp (\pi\alpha|1/r-r|) .
$$
 We assume that $s>0$. For a small perturbation ($\alpha <<1$),
the function on the RHS has a minimum  of
value $a_{min}(\alpha)\lesssim \left(\frac{\pi\alpha \e }{s}\right)^s$. This implies that
for any $n$, the spectrum
of $\P_{\eps|V_{n}}$ is contained in the disk
of radius $\bigl(a_{min}(\alpha)\bigr)^{|n|}$ for any $\eps>0$
(we have used the symmetry $a_{-n}=\overline{a_n}$).

For a more general perturbation 
$2\pi F'(t)=\sum _{m\geq 1}f_{m}t^m+c.c.$
one can use the bound $2\pi|\Im F'(rt)|\leq \sup \left( g(r),\, g(1/r)\right)$
with $g(r)\defi \sum _{m\geq 1}|f_{m}|r^{m}$. If $g(r)$ is ``small'',
then the function $r^{s}\exp (g(1/r))$ has a minimum $a_{min}<<1$ on $\R_+^*$. This
implies that the spectrum of $\P_{\eps|V_n}$ is contained, for any $\eps$, in the disk
around the origin of radius $a_{min}^{|n|}$.


\subsection{Quantum nonlinear shear\label{a:quantum nonlinear shear}}

The results of the previous appendix can be easily adapted to the quantized
nonlinear shear. The latter is defined as the composition of the quantum linear
shear $\hat{S}$ with the Floquet operator corresponding to the Hamiltonian
$\hat{F}$ (see Section \ref{s:nonlinear shear classical}): the
quantum map reads $\hat{S}\e^{-2\i \pi N\hat{F}}$. Its action
is diagonal in the impulsion basis $\{ |p_j\rangle_N\}$ of
$\hn $ ($p_j=\frac{j}{N}$, with $j\in \Z_N$):
$$
\hat{S}\e ^{-\i \hat{F}/\hbar }\, |p_{j}\rangle _{N}
=\e ^{-2\i \pi N\left( \frac{s}{2}p_{j}^{2}+F(p_{j})\right) }\, |p_{j}\rangle _{N}.
$$
Since $s$ is even, this expression is well-defined for any $j\in \Z_N$. As we now show,
it allows to express the action of the propagator $\hat{\P }$
on the quantum translations $\hat{T}_{k/N}$. 
Taking $k=(m,n)\in \Z_N^2$, this translation can be decomposed as 
$$
\hat{T}_{k/N}=\sum _{j\in \Z _{N}}|p_{j+n}\rangle \langle p_{j}|\, 
\e ^{-2\i \pi m(p_{j}+\frac{n}{2N})}.
$$
$\hat\P$ acts inside each subspace $V_{n,N}=Span\{\hat{T}_{(m,n)/N},\: m\in \Z_N\}=
Span\{|p_{j+n}\rangle \langle p_{j}|,\,j\in\Z_N\}$. The propagator $\hat\P$
indeed multiplies each $|p_{j+n}\rangle \langle p_{j}|$ by the phase 
$A_{n,N}\left( \e ^{2\i \pi (p_{j}+\frac{n}{2N})}\right)$, with the function
$A_{n,N}$ on $\T^1$ defined as:
$$
A_{n,N}(\e ^{2\i \pi p})\defi \e ^{-2\i \pi snp}
\e ^{-2\i \pi N[F(p+\frac{n}{2N})-F(p-\frac{n}{2N})]}
=\sum _{m\in \Z }\tilde{A}_{n,N}(m)\e ^{2\i \pi mp}.
$$
$A_{n,N}(t)$ uniformly converges to the function
$a_{n}(t)$ of Eq.~(\ref{e:function a_n}) in the limit $N\to \infty$.
For our example $F(p)=\frac{\alpha }{2\pi }\cos (2\pi p)$, it takes the concise form
$A_{n,N}(\e^{2\i \pi p})=\e^{-2\i\pi n(sp-\alpha_{n,N}\sin (2\pi p))}$
with $\alpha_{n,N}=\alpha \frac{N}{\pi n}\sin(\pi n/N)=\alpha(1+\O((n/N)^2))$. 

From there, we can express the action of $\hat\P$ on the
quantum translations:
$$
\hat{\P }\, \hat{T}_{\frac{(m,n)}{N}}=\sum _{l\in \Z }\hat{T}_{\frac{(m-l,n)}{N}}\: 
\tilde{A}_{n,N}(l)=\sum _{l\in \Z _{N}}\hat{T}_{\frac{(l,n)}{N}}\: \tilde{A}'_{n,N}(m-l),
$$
where $\tilde{A}'_{n,N}(m)=\sum_{\nu \in \Z }(-1)^{n\nu }\tilde{A}_{n,N}(m+\nu N)$.
Therefore, $\hat\P$ acts on $V_{n,N}$ through the $N\times N$
Toeplitz matrix with coefficients $\tilde{A}'_{n,N}(m)$.
If we truncate this matrix to the size $\{|l|,|l'|\leq \eps^{-1}\}$ with $\eps^{-1}<<N$, 
we only take into account the coefficients with $|m|\leq 2\eps^{-1}$.
Since $A_{n,N}$ is a smooth function,
each of these coefficients is the sum of a ``dominant'' term $\tilde{A}_{n,N}(m)$
and a ``remainder''. 
For our example, the classical coefficients are given by Bessel functions:
$\tilde{a}_n(m)=J_{m+sn}(2\pi n\alpha)$, and the same for $\tilde{A}_{n,N}(m)$,
replacing $\alpha$ by $\alpha_{n,N}$. These coefficients therefore satisfy
$|\tilde{A}_{n,N}(m)|\leq C\frac{(n\alpha)^{|m+ns|}}{|m+ns|!}$,
so that the remainder is uniformly bounded above by 
$\left(\frac{C' \alpha }{\eps N}\right)^{N/2}$
for large $N$. If we call $\hat\P^\sharp_{\eps|V_{n,N}}$ the truncated Toeplitz matrix with
coefficients $\tilde{A}_{n,N}(m)$, we get the estimate:
$$
\Vert \hat{\P}_{\eps|V_{n,N}} - \hat\P^\sharp_{\eps|V_{n,N}} \Vert 
\leq (2\eps^{-1}+1)\left(\frac{C'\alpha}{\eps N}\right)^{N/2}.
$$  
This estimate implies that the spectra of
both matrices cannot be at a distance larger than $\O((\eps N)^{-\eps N/4})$.  
The matrix $\hat\P^\sharp_{\eps|V_{n,N}}$ may be analyzed along the same lines
as $\P_{\eps|V_{n}}$ in the previous Appendix. Its spectrum
is contained for any $\eps>0$ and any $r>0$ in the convex hull of $A_{n,N}(r\T^1)$, 
which converges to the convex hull of 
$a_n(r\T^1)$ when $N\to\infty$.

\end{document}